\documentclass[aps,pre,twocolumn,superscriptaddress,showpacs,reprint]{revtex4-1}
\usepackage{graphicx}
\usepackage{latexsym}
\usepackage{amssymb}
\usepackage{mathrsfs}
\usepackage{amsmath}
\usepackage{soul}
\usepackage{bm}
\usepackage{verbatim}
\usepackage{color}
\usepackage{xcolor}
\usepackage{hyperref}
\usepackage{epsfig}
\usepackage{multirow}

\DeclareMathOperator{\sign}{sign}

\newcommand{\pvec}[1]{\vec{#1}\mkern2mu\vphantom{#1}}

\begin{document}

\title{{\normalsize{}Theory for the density of interacting quasi-localised modes in amorphous solids}}
\author{Wencheng Ji}
\affiliation{Institute of Physics, EPFL, CH-1015 Lausanne, Switzerland}
\author{Marko Popovi{\' c}}
\affiliation{Institute of Physics, EPFL, CH-1015 Lausanne, Switzerland}
\author{Tom W. J. de Geus}
\affiliation{Institute of Physics, EPFL, CH-1015 Lausanne, Switzerland}

\author{ Edan Lerner}
\affiliation{Institute for Theoretical Physics, University of Amsterdam, Science Park 904, 1098 XH Amsterdam, The Netherlands}
\author {Matthieu Wyart}
\affiliation{Institute of Physics, EPFL, CH-1015 Lausanne, Switzerland}

\date{\today}
\begin{abstract}
Quasi-localised modes appear in the vibrational  spectrum of amorphous solids at low-frequency. Though never formalised, these modes are believed to have a close relationship with other important local excitations, including shear transformations and two-level systems. We provide a theory for their frequency density, $D_{L}(\omega)\sim\omega^{\alpha}$, that establishes this link for systems at zero temperature under quasi-static loading.
It predicts two regimes depending on the density of shear transformations $P(x)\sim x^{\theta}$ (with $x$ the additional stress needed to trigger a shear transformation). If $\theta>1/4$, $\alpha=4$ and a finite fraction of quasi-localised modes form shear transformations, whose amplitudes vanish at low frequencies. If $\theta<1/4$, $\alpha=3+ 4 \theta$ and all quasi-localised modes form shear transformations with a finite amplitude at vanishing frequencies. We confirm our predictions numerically.
\end{abstract}

\maketitle

\section{Introduction}

Unlike crystals, amorphous solids do not present topological defects due to their lack of long-range order. Instead they display excitations where a group of particles can rearrange.
These essentially local excitations lead to a dipolar change of stress in the medium, which can effectively couple them. An example of local excitations is \emph{two-level systems}
that  govern the low-temperature properties of glasses, for which the particles' rearrangement is induced by quantum tunnelling \cite{Anderson72,Phillips72,Phillips81}.
The nature of two-level systems and the role of their interactions, argued to lead to a pseudo-gap in their density, is still debated \cite{Yu88,Faoro15,Parshin07}.
Another example of local excitations is \emph{shear transformations} \cite{Argon79,Falk98,Tanguy06,Schall07,Amon12}, in which a rearrangement, or plastic event, can occur in the absence of any quantum or thermal fluctuations by a local increase in stress that triggers a saddle node bifurcation \cite{Maloney06a}. In this case, the important role of interactions is established: they lead to bursts of avalanches of many plastic events \cite{Lemaitre09,Maloney09}. This behaviour is a necessary consequence \cite{Muller14,Lin15} of the presence of a pseudo-gap $P(x)\sim x^\theta$
in the density of these excitations \cite{Lemaitre07,Karmakar10a, Lin14,Lin14a,Budrikis17} (where $x$ is the additional shear stress that must be applied locally to trigger a new event). Treating the effect of interactions as a mean-field mechanical noise leads to the prediction that $\theta$ varies non-monotonically as shear stress is applied \cite{Lin16}, as confirmed numerically \cite{Lin15,Hentschel15,Ozawa18}.

These local excitations correspond to directions in phase space with little restoring forces, suggesting that the low-frequency part of the vibrational spectrum contains information on their respective nature.
This view is supported by the early observation that quasi-localised modes are present at low-frequencies \cite{Schober96},
leading to a considerable numerical  effort to characterise them. Most studies find that their density follows
$D_{L}(\omega)\sim\omega^{\alpha}$ with $\alpha=4$ \cite{Ilyin87,Baity15,Gartner16,Lerner16,Mizuno17,Stanifer18,Wang18} although $\alpha\approx 3$ has also been reported \cite{Lerner17,Xu17}. Theoretically, it has been argued, for general bosonic disordered systems, that $\alpha=4$ in the ground state, and $\alpha=3$ in generic meta-stable states \cite{Gurevich03,Gurarie03}.
This theory, however, neglects  interactions between quasi-localised modes.
Its apparent success thus seems to be at odds with the established role of interactions in determining the properties of plastic deformation and yielding \cite{Lemaitre09,Maloney09,Muller14,Lin15, Lemaitre07,Karmakar10a,Lin14,Lin14a,Budrikis17}.

In this article, we provide a theory for the density of quasi-localised modes for classical systems at zero temperature,
which takes their interactions into account and clarifies their relationship with shear transformations. In particular, we introduce and treat analytically
a mesoscopic model of interacting quasi-localised modes. We predict two distinct regimes, shear transformations are found to be the dominant source of quasi-localised modes only in one of them.
 We confirm our predictions by independently measuring the exponents $\alpha$ and $\theta$, using molecular dynamics simulations of quasi-statically sheared glasses obtained at distinctly different quench rates.

\section{Mesoscopic model \& theoretical prediction}

 We model an amorphous solid as a collection of mesoscopic blocks whose size is comparable to that of quasi-localised modes. In each block we consider the softest quasi-localised mode. We denote by $s$ the displacement along that mode and by $u_i(s)$ the Taylor expansion of the energy \cite{Karpov83} in a block $i$:
\begin{align}
u_i(s)=\frac{1}{2!}\lambda_i s^{2}+\frac{1}{3!}\kappa_i s^{3}+\frac{1}{4!}\chi_i s^{4} + \mathcal{O}(s^5) . \label{eq:1}
\end{align}
Numerical measurements of $\chi_i$ have shown that its distribution is narrow \cite{Lerner16}, we thus assume that it does not depend on $i$, and choose the units of the displacement $s$  so that  $\chi=1$. Consequently $\lambda_i$ and $\kappa_i$ determine the shape the potential. Note that $\lambda_i = \omega^2_i$ is the smallest eigenvalue of the Hessian of the block.

\begin{figure}[ht]
  \includegraphics[width=.95\linewidth]{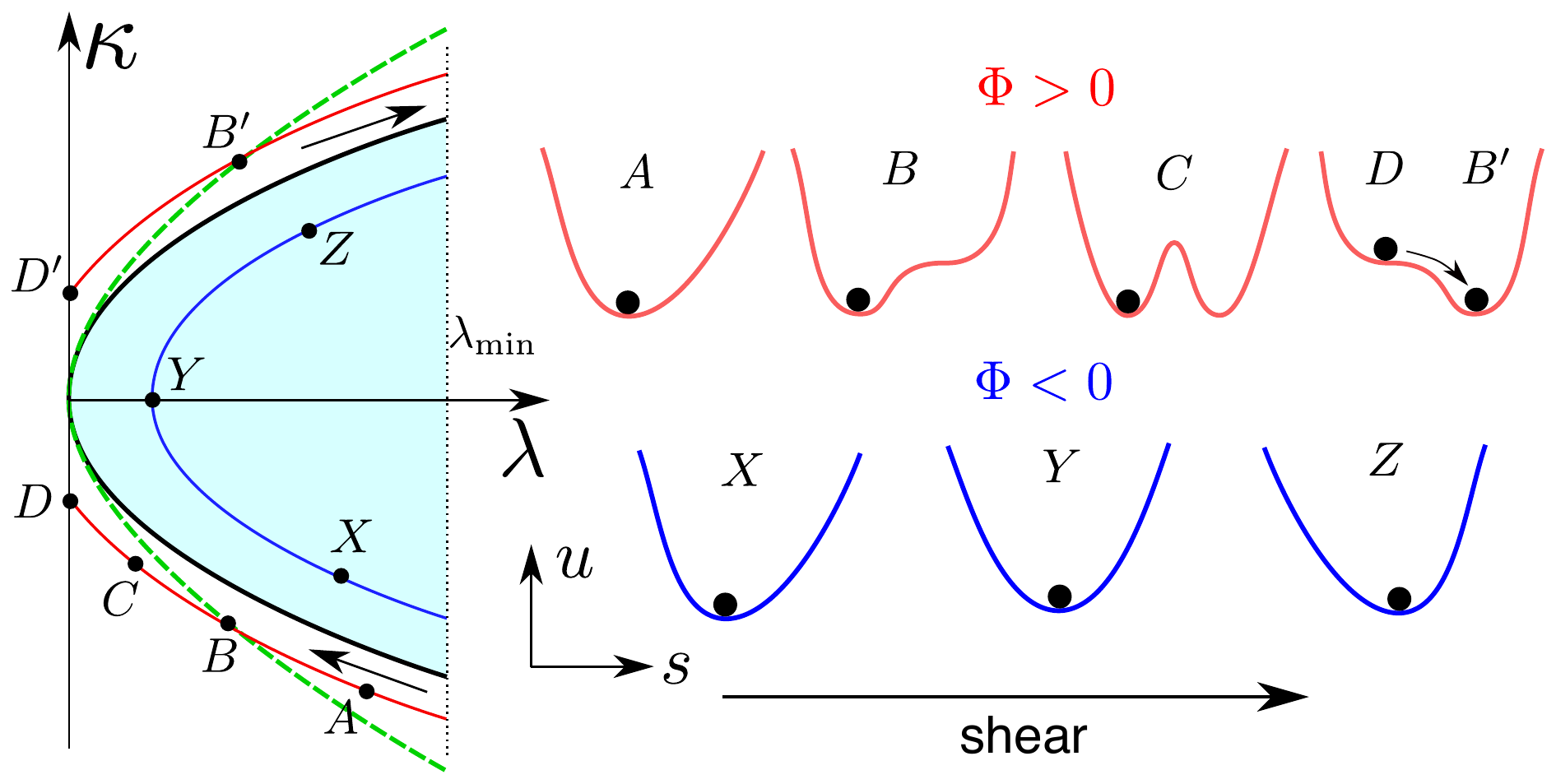}
  \caption{Illustration of the flow of block parameters in the $\left(\lambda, \kappa\right)$ plane, under increasing shear stress. Examples of trajectories are shown on the left. The shape of the associated potentials $u$ are shown on the right. The blue line is an example of a passive mode, for which $\Phi<0$. The red line is an example of a shear transformation, for which $\Phi>0$. The black line corresponds to the marginal case for which $\Phi=0$. The dashed green line marks the appearance of the second minimum in $u$, and thus the locations of reinsertion after a failure.}
  \label{fig1}
\end{figure}

The shear stress $\sigma_i$ in the block can change either due to a global applied stress or due to an interaction with another block in which a rearrangement occurred. A change of shear stress by $\delta \sigma_i$ tilts the potential $u_i(s)$:
\begin{align}
\tilde{u}_i(s,\; \delta \sigma_i)=u_i(s)-s \; C_i\delta \sigma_i,\label{eq:2}
\end{align}
In this scalar approximation, $C_i$ describes the coupling between this mode and the shear stress, and should depend on $i$, and possibly on the value of local stress $\sigma_i$.
We neglect these dependencies and impose $C_i=1$ through a suitable choice of the units of stress.
Following ideas presented in \cite{Karpov85} we expand the energy around the new minimum. This changes $\lambda_i$ and $\kappa_i$, and in the limit of infinitesimal $\delta \sigma_i$ we obtain the following flow:
\begin{equation}
\frac{\partial\lambda_i}{\partial \sigma_i}  =\frac{\kappa_i}{\lambda_i},\label{eq:3} \ \ \ \ \ \ \ \ \frac{\partial \kappa_i}{\partial \sigma_i}  =\frac{1}{\lambda_i}.
\end{equation}
See Appendix~\ref{App.A} for details.
A conserved quantity of this dynamics is:
\begin{align}
\Phi_i\equiv \kappa_i^{2}-2\lambda_i. \label{eq:5}
\end{align}
Thus, we can track the evolution of the energy shape in each block along the parabolic trajectories in the $(\lambda,\kappa)$ plane.
As we illustrate in Fig. \ref{fig1}, two distinct behaviours, as previously identified in \cite{Karpov85}, are separated by the $\Phi= 0$ parabola (black line).
 $\Phi < 0$ (highlighted in blue) corresponds to `passive' modes that never undergo a saddle node bifurcation. For $\Phi>0$ (in red) shear transformations occur. In that case a potential with a single minimum (point $A$) evolves under increasing stress to a point where a second minimum appears (point $B$). As the stress increases, the minima become equally deep (point $C$). Eventually, a saddle node bifurcation occurs (point $D$) and the system falls in the other minimum (point $B'$). It is straightforward to show that points $B$ and $B'$ lie on the parabola $\kappa^{2}-8\lambda/3=0$ (see Appendix~\ref{App.B}), indicated using a dashed green line in Fig.~\ref{fig1}.

So far we have described how to track the softest quasi-localised mode in each block. However, it may happen that the tracked mode becomes stiffer than the next softest mode. This will occur for a typical value of $\lambda$ that we denote $\lambda_{\text{min}}$. To implement this effect in the model, if $\lambda$ reaches $\lambda_{\min}$ we switch to a new softest mode. We describe its property by choosing its $\Phi$ randomly from a distribution $P_0(\Phi)$. We expect that $P_0(\Phi)$ is a smooth distribution, i.e.\ that $P_0(\Phi)>0$ in the relevant range of $\Phi$ (without any singularity, in particular around $\Phi=0$).

Following elasto-plastic models \cite{Picard04,Baret02,Nicolas17}, we describe the change of shear stress on block $i$ as $\delta \sigma_i=d\Sigma +\eta_i$, where $d\Sigma$ is the increment of  globally applied  stress, and $\eta_i$ stems from the rearrangements (saddle node bifurcations) of other blocks. $\eta_i$ is of zero mean and displays a power-law distribution \cite{Lemaitre07,Lin16}.
During an avalanche of rearrangements $\delta \sigma_i$ thus performs a random flight. In the mean-field approximation where $\eta_i$ is assumed to be uncorrelated in space and time, it corresponds to a L\'{e}vy Flight, and exact calculations are possible \cite{Lin16}. Our arguments below, however, do not rely on this approximation.

The spectrum of the Hessian, $P(\lambda)$, can be calculated as a marginal distribution of the density of states $P(\lambda, \kappa)$. A change of variables allows us to express $P(\lambda, \kappa)$ in terms of $\Delta \sigma$ and $\Phi$:
\begin{align}
P(\lambda,\kappa)=2\lambda \; P(\Delta \sigma,\Phi),  \label{eq:6}
\end{align}
where $\Delta \sigma$ is the accumulated stress change relative to an arbitrary reference and the factor $2\lambda$ corresponds to the absolute value of the Jacobian $|\partial(\Delta \sigma, \Phi)/\partial(\lambda, \kappa)|$ (see Appendix~\ref{App.C}).
We first consider passive modes for which $\Phi<0$, and for convenience chose to define $\Delta\sigma$ such that $\Delta\sigma=0$
at $\kappa=0$.  If many rearrangements take place in the system (as expected after a fast quench or after a succession of avalanches
triggered by increasing the stress), then the flights in each block will lead to a finite distribution for $P(\Delta\sigma)$ at any $\Phi<0$, independently
of the initial conditions, as long as $P_0(\Phi)>0$. In particular  $P(\Delta\sigma=0,\Phi=0)>0$, implying that $P(\lambda,\kappa)\sim \lambda$ in the limit of
vanishing $\lambda$. Thus the contribution of passive modes to the spectrum of the Hessian is:
\begin{align}
\label{eq:8}
P(\lambda)= \int_{\Phi \leq 0}P(\lambda,\kappa)d\kappa\sim \lambda\int_{\kappa \leq \sqrt{2\lambda}}  d\kappa \sim \lambda^{3/2} .
\end{align}

For a fixed $\Phi > 0$, failure occurs when  $\lambda= 0$. An example is shown in Fig.~\ref{fig1} as point $D$ and its mirror image $D'$. Thus the dynamics after a fast quench or a big avalanche corresponds to a stochastic walk with absorbing conditions at these points, and reinsertion in points $B'$ ($B$) if failure happened in $D$ ($D'$).

We then identify $x = \Delta\sigma - \left.\Delta\sigma\right|_{\lambda= 0}$ as the additional stress needed to trigger a shear transformation. The density of states can be shown to display a power-law between the absorption and reinsertion points $D$ and $B$ \footnote{Blocks are also inserted at $\lambda = \lambda_{\text{min}}$ as discussed earlier. However, as long as a finite fraction of blocks are reinserted they give a dominant contribution to the density of states. For example, in a mean-field approximation blocks perform L\'{e}vy Flight of index $\mu = 1$ \cite{Lin16} in which diffusion and drift are comparable so that a finite fraction of blocks fails in opposite direction of the drift and is reinserted.}. From \cite{Lin16} we know that $P(x,\Phi)\sim (x/x^*)^\theta$ for $x<x^*$, where $x^*$ corresponds to point $B$. For $x\gg x^*$, $P(x)$ will vary smoothly. It is straightforward to show that $x^*\sim \Phi^{3/2}$
and that for $x \ll x^*$, $x\sim \lambda^2/\kappa$ and $\Phi\sim \kappa^2$ (see Appendix~\ref{App.D}). Using Eq.~\eqref{eq:6} we finally obtain:
\begin{equation}
\label{eq:16}
 P(\lambda, \kappa) \sim \frac{ \lambda^{2\theta + 1}}{\kappa^{4\theta}}  \ \ \ \hbox{ for } \  \lambda \ll \kappa^2.
\end{equation}
Eq.~\eqref{eq:16} readily gives the contribution of shear transformations to the spectrum of the Hessian:
\begin{align}
\label{eq:7}
 P(\lambda)= \int_{\Phi > 0}P(\lambda,\kappa)d\kappa \sim
  \begin{cases}
    \lambda^{2\theta + 1} & \theta < 1/4  \\
    \lambda^{3/2} & \theta \geq 1/4
    \end{cases}
    .
\end{align}
Thus if $\theta<1/4$, shear transformations dominate the low-frequency spectrum of the Hessian. Following Eq.~\eqref{eq:16},  the integral in Eq.~\eqref{eq:7} is dominated by  large $\kappa$, implying
that shear transformations leading to large plastic events are observed as $\lambda\rightarrow 0$. By contrast, for $\theta\geq 1/4$ both shear transformations and passive modes contribute equally. In that case the integral in Eq.~\eqref{eq:7} is dominated by  small $\kappa\sim \sqrt{\lambda}$, implying that  low-frequency shear transformations lead to tiny rearrangements.  Concerning the density of vibrational modes, using $\omega^2\sim \lambda$ and Eqs.~(\ref{eq:8},\ref{eq:7}) we get:
\begin{align}
  D_L(\omega)= P(\lambda) \frac{d\lambda}{d\omega} \sim \omega^{\alpha}, \quad
  \alpha= \begin{cases}
	4\theta+3 & \theta<1/4\\
	4  & \theta\geq1/4
\end{cases}
 .\label{eq:13}
\end{align}

Note that in the absence of interactions $\theta = 0$, and consequently our result is consistent with the theoretical prediction for non-interacting modes $\alpha =3$ \cite{Gurevich03,Gurarie03}. In the presence of interactions $\theta$ will generally be nonzero and will depend on the system preparation. The latter allows us to test our theory, which we now do using molecular dynamics simulations for different preparation protocols.

%

\section{Molecular dynamics}
As mentioned in the introduction, mean-field theory predicts that $\theta=1/2$ after a quench, then drops and rises again as a function of the applied shear strain $\gamma$ \cite{Lin16}. The drop is expected to be more pronounced for well-annealed glasses \cite{Lin16}, as confirmed numerically \cite{Lin15,Hentschel15,Ozawa18}. To test our theory we thus measure $\alpha$ and $\theta$ as a function of $\gamma$ for glasses obtained using different preparations.
We consider the three-dimensional bi-disperse glass of \cite{Lerner17}, composed of $N$ particles interacting by purely repulsive inverse power-law potentials, which are continuous up to the third derivatives. We consider two distinct preparation protocols: (\emph{i}) a \emph{rapid quench}, that results in a poorly annealed glass, obtained by a steepest descent after instantaneously cooling from a temperature $T = T_0$ (higher than the glass transition temperature) to $T = 0$; and (\emph{ii}) a \emph{slow quench}, that results in a better annealed glass, obtained by first cooling it at a low rate from $T = T_0$ to $T = T_0/10$, followed by a steepest descent to remove the remaining heat. Details are provided in the Appendix~\ref{App.E}.

\begin{figure}[ht]
  \includegraphics[width=.95\linewidth]{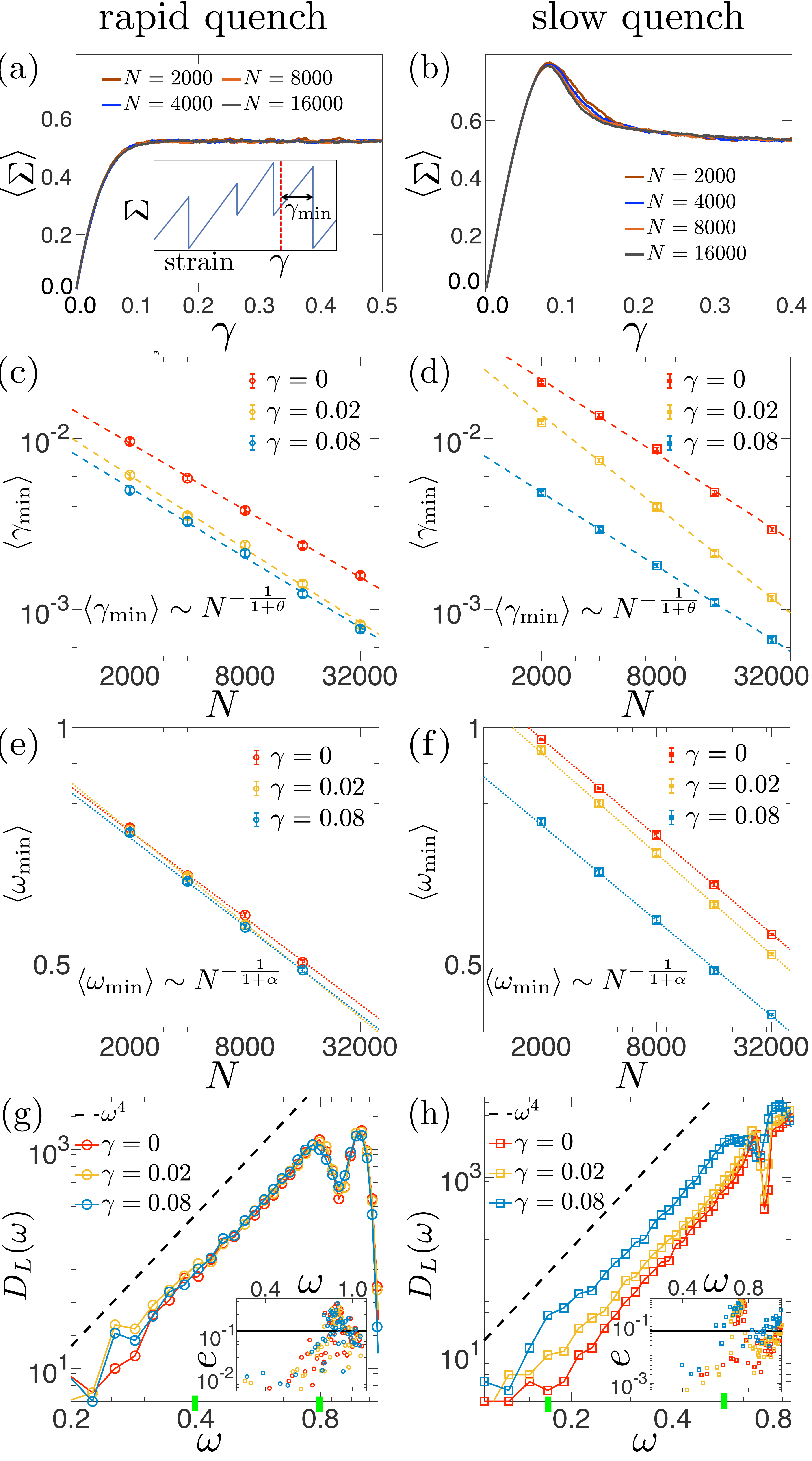}
  \caption{(a,b) Stress-strain curves averaged over 1000 realisations for different $N$ (see legend).
  (c,d) Finite size scaling of the mean strain increment between plastic events $\langle\gamma_{\min}\rangle$ at three representative applied strains to determine $\theta$ (see legend).  (e,f) Finite size scaling of the mean lowest frequency is used at three representative  strains $\gamma$ to determine $\alpha$ (see legend).
  (g,h) Density of quasi-localised modes $D_L(\omega)$ at low frequencies $\omega$: (g) 1000 realisations at $N=16000$ and (h) 5000 realisations at $N=32000$. Modes with a participation ratio above the threshold (g) $e_{c}=0.125$ and (h) $e_{c}=0.0625$ have been removed following the procedure of \cite{Mizuno17}. Insets of (g,h) show participation ratios $e$ and their thresholds $e_c$ (solid black lines). The green markers on the axes of (g,h) indicate the fitting range of $\omega$.
  }
  \label{fig2}
\end{figure}

\begin{figure}[ht]
  \includegraphics[width=.95\linewidth]{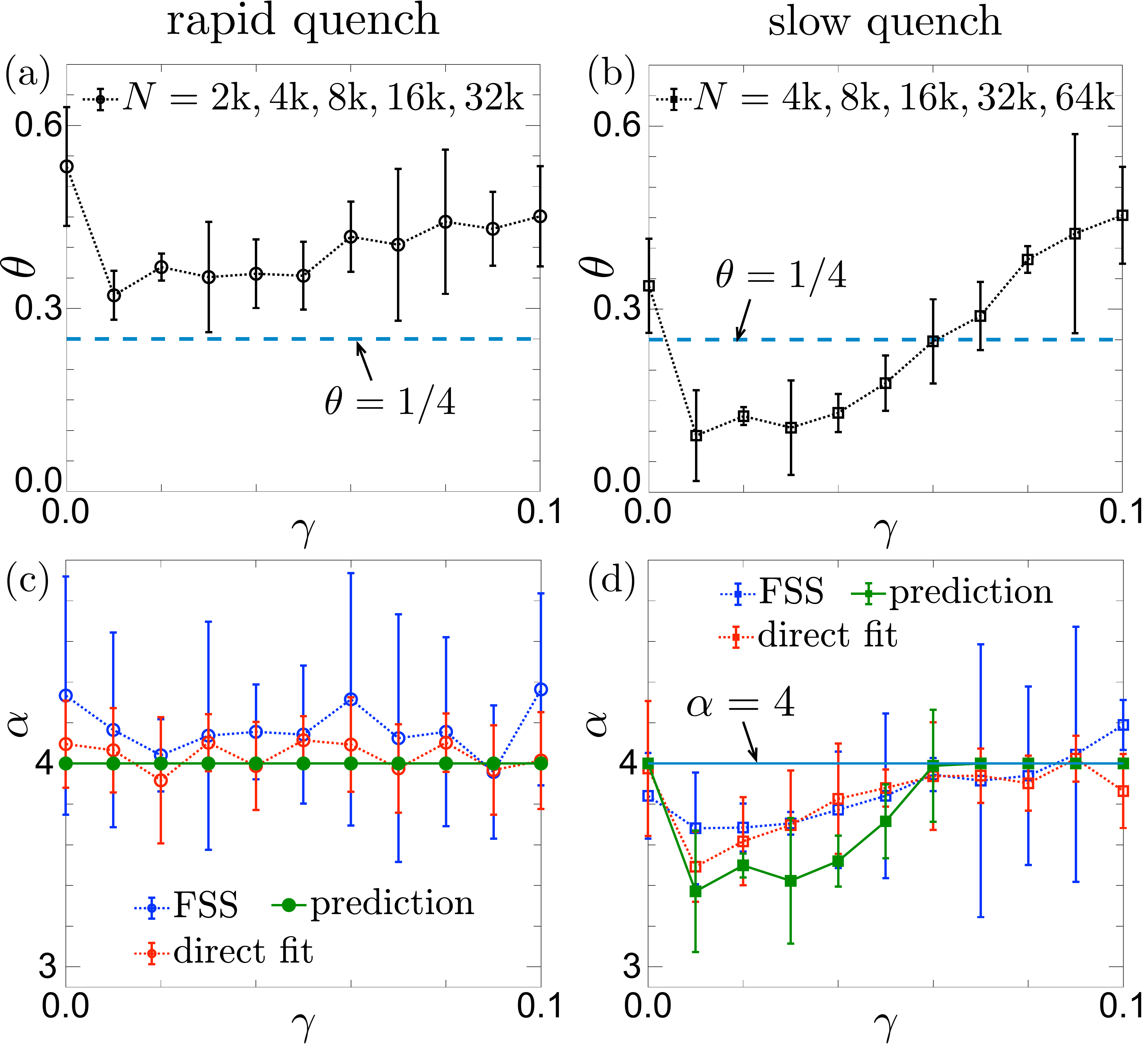}
  \caption{(a,b): Exponent $\theta$ extracted by finite size scaling of $\langle \gamma_{\min}\rangle$ at different $\gamma$. The employed system sizes are reported in the legends. (c,d): Green line: Prediction of $\alpha$ based on measured $\theta$. Blue line: $\alpha$ extracted by finite size scaling (FSS) of $\langle \omega_{\min}\rangle$ at different $\gamma$ using system size range $N=2000,4000,8000,16000$ and $N=4000,8000,16000,32000$. Red line: $\alpha$ obtained from a direct fit $D_L(\omega)$ at $N=16000$ and $N=32000$. Our prediction for $\alpha$ is indicated using a solid green line. The error bars for $\theta$ and $\alpha$ correspond to 95\% confidence intervals for the coefficient estimates obtained by linear fit on a log-log scale.}
  \label{fig3}
\end{figure}

After the glass is prepared, we quasi-statically apply a simple shear using Lees-Edwards periodic boundary conditions \cite{Allen91}.
As commonly reported, we find that the stress-strain curve  $\langle \Sigma(\gamma) \rangle$ is monotonic after a rapid quench and displays an overshoot after a slower quench,
as shown in Figs.~\ref{fig2}(a,b). The pseudo-gap exponent $\theta$ is readily extracted using extreme value statistics \cite{Karmakar10a,Lin14a}, which uses the fact that $\langle x_{\min}\rangle \sim N^{-1/(1+\theta)}$, where $x_{\min}$ characterises the shear transformation closest to an instability. More precisely, it is the additional stress needed to trigger the next plastic event. It is proportional to the strain increment between events, $\gamma_{\min}$, illustrated in the inset of Fig.~\ref{fig2}(a). $\langle \gamma_{\min}(N)\rangle$ is reported in Figs.~\ref{fig2}(c,d), from which the exponent $\theta$ is extracted via a power-law fit. The result is reported in Fig.~\ref{fig3}(a,b) where the predicted non-monotonicity of $\theta(\gamma)$ is observed. We find that for the rapidly quenched glass, $\theta>1/4$ for all $\gamma$, leading to the prediction that $\alpha=4$. By contrast, the slowly quenched glass displays a range of strains for which $\theta<1/4$, where our prediction is that $\alpha < 4$.

To measure the exponent $\alpha$, we diagonalise the Hessian to obtain $D_L(\omega)$. We then determine $\alpha$ in two ways. One way is to use the fact that the mean lowest frequency scales with the system size as $\langle \omega_{\min}\rangle \sim N^{-1/(1 + \alpha)}$. We show this scaling at three representative values of strain $\gamma$ in Figs.~\ref{fig2}(e,f) and measured values of $\alpha$ are shown as blue line in Figs.~\ref{fig3}(c,d).

The exponent $\alpha$ can also be measured directly from $D_L(\omega)$, in contrast to $\theta$, that cannot be obtained directly from the distribution $P(\gamma_{\min})$, as we explain in  Appendix~\ref{App.F}. However, this measurement is challenging as it is polluted by the influence of plane waves and by finite size effects at low frequencies (see Figs.~\ref{fig2}(g,h) and Appendix~\ref{App.H}). In order to perform this measurement, we follow a protocol \cite{Mizuno17} that separates quasi-localised modes from plane waves based on their participation ratio $e \equiv 1/ ( N \; \sum_{j}(\mathbf{\Psi}_{j}^{2})^{2} )$, where $\mathbf{\Psi}_{j}$ is the eigenmode component on the $j$th particle. Examples of participation ratios are shown in insets of Figs.~\ref{fig2}(g,h) where the employed threshold $e_c$ is indicated by a horizontal line. We have verified robustness of our results below by raising and lowering $e_c$ by $20\%$. $\alpha$ is fit on the `filtered' $D_L(\omega)$, whereby the fitting range is bounded on the upper side by the point where the power law scaling is clearly interrupted by the plane waves. A range of lower bounds has been used for which the measurement of $\alpha$ is robust (see Appendix~\ref{App.G} for details). The employed fitting range is indicated using green bars in Figs.~\ref{fig2}(g,h).

The results, in Fig.~\ref{fig3}(c,d), show that the two different measurements of $\alpha$ are in a good qualitative agreement. The results are consistent with our theoretical prediction for $\alpha$. In rapidly quenched systems $\alpha = 4$, also true in the steady state (shown in Appendix~\ref{App.I}), while in slowly quenched systems we find that $\alpha$ is significantly smaller than $4$ precisely in the range where $\theta<1/4$. To our knowledge, this is the first time that such a non-monotonic behaviour of $\alpha$ as a function of shear has been measured directly from MD simulations.

\section{Conclusion}

We have provided a theory for the density of localised soft modes in classical amorphous solids at zero temperature. 

Our approach goes beyond previous ones by taking long-range interactions between these modes into account. 
We have found two regimes, one in which modes near a saddle-node bifurcation are dominant, and one in which they contribute to a finite fraction of the spectrum (the rest consisting of passive modes that are irrelevant as far as plasticity and two-level systems are concerned). The first regime does not appear in rapidly quenched materials (and is thus presumably absent in foams and granular materials). By contrast it is expected to be very pronounced in real glasses which are much more stable than the `slowly quenched' configurations studied here. This view is supported by recent measurements in simulated glasses prepared by a swap algorithm (that are comparable to experimental cooling rates), which indeed show extremely small values for $\theta$ \cite{Ozawa18}. Note that our argument appears to be rather generic, and may apply to other disordered systems with long-range interactions, e.g.\ in crystals with defects.

Our work is a necessary first step to describe systems at finite temperatures or shear rates. For example, it is interesting to reflect on the role of thermal fluctuations in a perturbative manner, if a very small temperature would have been switched on in the configurations we have described. Modes in which the high energy well is occupied would eventually switch states (an effect that is faster for small $\lambda $ and $\kappa$ where barriers are small and activation is fast). In the $(\lambda,\kappa)$ plane this would lead to a depleted region, whose right border corresponds to a limiting parabola where wells are of equal depth (including point $C$ and the origin Fig.~\ref{fig1}). In time the depleted region will grow away from the origin and a pseudo-gap may open at the limiting parabola \cite{Efros75, Faoro15,Kapteijns18}.
Away from this region we expect our described solutions to hold. Measuring the joint distribution $P(\lambda,\kappa)$ for this kind of protocol, a task for which recent numerical methods are being designed \cite{Lerner17}, would shed light on the nature of bottom of the energy landscape in glasses.

 \begin{acknowledgments}
We thank  A. Rosso, J. Lin and the Simons collaboration for discussions. M.~W.~thanks the Swiss National Science Foundation for support under Grant No. 200021-165509 and the Simons Foundation Grant ($\#$454953 Matthieu Wyart).
T.~G.~was partly financially supported by The Netherlands Organisation for Scientific Research (NWO) by a NWO Rubicon grant number 680-50-1520.
\end{acknowledgments}
\bibliographystyle{unsrt}

\bibliographystyle{apsrev4-1}
\bibliography{Wyartbibnew}

\begin{thebibliography}{45}%
\makeatletter
\providecommand \@ifxundefined [1]{%
 \@ifx{#1\undefined}
}%
\providecommand \@ifnum [1]{%
 \ifnum #1\expandafter \@firstoftwo
 \else \expandafter \@secondoftwo
 \fi
}%
\providecommand \@ifx [1]{%
 \ifx #1\expandafter \@firstoftwo
 \else \expandafter \@secondoftwo
 \fi
}%
\providecommand \natexlab [1]{#1}%
\providecommand \enquote  [1]{``#1''}%
\providecommand \bibnamefont  [1]{#1}%
\providecommand \bibfnamefont [1]{#1}%
\providecommand \citenamefont [1]{#1}%
\providecommand \href@noop [0]{\@secondoftwo}%
\providecommand \href [0]{\begingroup \@sanitize@url \@href}%
\providecommand \@href[1]{\@@startlink{#1}\@@href}%
\providecommand \@@href[1]{\endgroup#1\@@endlink}%
\providecommand \@sanitize@url [0]{\catcode `\\12\catcode `\$12\catcode
  `\&12\catcode `\#12\catcode `\^12\catcode `\_12\catcode `\%12\relax}%
\providecommand \@@startlink[1]{}%
\providecommand \@@endlink[0]{}%
\providecommand \url  [0]{\begingroup\@sanitize@url \@url }%
\providecommand \@url [1]{\endgroup\@href {#1}{\urlprefix }}%
\providecommand \urlprefix  [0]{URL }%
\providecommand \Eprint [0]{\href }%
\providecommand \doibase [0]{http://dx.doi.org/}%
\providecommand \selectlanguage [0]{\@gobble}%
\providecommand \bibinfo  [0]{\@secondoftwo}%
\providecommand \bibfield  [0]{\@secondoftwo}%
\providecommand \translation [1]{[#1]}%
\providecommand \BibitemOpen [0]{}%
\providecommand \bibitemStop [0]{}%
\providecommand \bibitemNoStop [0]{.\EOS\space}%
\providecommand \EOS [0]{\spacefactor3000\relax}%
\providecommand \BibitemShut  [1]{\csname bibitem#1\endcsname}%
\let\auto@bib@innerbib\@empty
\bibitem [{\citenamefont {Anderson}\ \emph {et~al.}(1972)\citenamefont
  {Anderson}, \citenamefont {Halperin},\ and\ \citenamefont
  {Varma}}]{Anderson72}%
  \BibitemOpen
  \bibfield  {author} {\bibinfo {author} {\bibfnamefont {P.~W.}\ \bibnamefont
  {Anderson}}, \bibinfo {author} {\bibfnamefont {B.~I.}\ \bibnamefont
  {Halperin}}, \ and\ \bibinfo {author} {\bibfnamefont {C.~M.}\ \bibnamefont
  {Varma}},\ }\href@noop {} {\bibfield  {journal} {\bibinfo  {journal} {Philos.
  Mag.}\ }\textbf {\bibinfo {volume} {25}},\ \bibinfo {pages} {1} (\bibinfo
  {year} {1972})}\BibitemShut {NoStop}%
\bibitem [{\citenamefont {Phillips}(1972)}]{Phillips72}%
  \BibitemOpen
  \bibfield  {author} {\bibinfo {author} {\bibfnamefont {W.}~\bibnamefont
  {Phillips}},\ }\href@noop {} {\bibfield  {journal} {\bibinfo  {journal}
  {Journal of Low Temperature Physics}\ }\textbf {\bibinfo {volume} {7}},\
  \bibinfo {pages} {351} (\bibinfo {year} {1972})}\BibitemShut {NoStop}%
\bibitem [{\citenamefont {Anderson}(1981)}]{Phillips81}%
  \BibitemOpen
  \bibfield  {author} {\bibinfo {author} {\bibfnamefont {A.}~\bibnamefont
  {Anderson}},\ }\href@noop {} {\emph {\bibinfo {title} {Amorphous Solids: Low
  Temperature Properties}}},\ edited by\ \bibinfo {editor} {\bibfnamefont
  {W.~A.}\ \bibnamefont {Phillips}},\ \bibinfo {series} {Topics in Current
  Physics}, Vol.~\bibinfo {volume} {24}\ (\bibinfo  {publisher} {Springer,
  Berlin},\ \bibinfo {year} {1981})\BibitemShut {NoStop}%
\bibitem [{\citenamefont {Yu}\ and\ \citenamefont {Leggett}(1988)}]{Yu88}%
  \BibitemOpen
  \bibfield  {author} {\bibinfo {author} {\bibfnamefont {C.}~\bibnamefont
  {Yu}}\ and\ \bibinfo {author} {\bibfnamefont {A.}~\bibnamefont {Leggett}},\
  }\href@noop {} {\bibfield  {journal} {\bibinfo  {journal} {Comments on
  Condensed Matter Physics}\ }\textbf {\bibinfo {volume} {14}},\ \bibinfo
  {pages} {231} (\bibinfo {year} {1988})}\BibitemShut {NoStop}%
\bibitem [{\citenamefont {Faoro}\ and\ \citenamefont {Ioffe}(2015)}]{Faoro15}%
  \BibitemOpen
  \bibfield  {author} {\bibinfo {author} {\bibfnamefont {L.}~\bibnamefont
  {Faoro}}\ and\ \bibinfo {author} {\bibfnamefont {L.~B.}\ \bibnamefont
  {Ioffe}},\ }\href@noop {} {\bibfield  {journal} {\bibinfo  {journal} {Phys.\
  Rev.\ B}\ }\textbf {\bibinfo {volume} {91}},\ \bibinfo {pages} {014201}
  (\bibinfo {year} {2015})}\BibitemShut {NoStop}%
\bibitem [{\citenamefont {Parshin}\ \emph {et~al.}(2007)\citenamefont
  {Parshin}, \citenamefont {Schober},\ and\ \citenamefont
  {Gurevich}}]{Parshin07}%
  \BibitemOpen
  \bibfield  {author} {\bibinfo {author} {\bibfnamefont {D.}~\bibnamefont
  {Parshin}}, \bibinfo {author} {\bibfnamefont {H.}~\bibnamefont {Schober}}, \
  and\ \bibinfo {author} {\bibfnamefont {V.}~\bibnamefont {Gurevich}},\
  }\href@noop {} {\bibfield  {journal} {\bibinfo  {journal} {Phys.\ Rev.\ B}\
  }\textbf {\bibinfo {volume} {76}},\ \bibinfo {pages} {064206} (\bibinfo
  {year} {2007})}\BibitemShut {NoStop}%
\bibitem [{\citenamefont {Argon}(1979)}]{Argon79}%
  \BibitemOpen
  \bibfield  {author} {\bibinfo {author} {\bibfnamefont {A.}~\bibnamefont
  {Argon}},\ }\href {\doibase 10.1016/0001-6160(79)90055-5} {\bibfield
  {journal} {\bibinfo  {journal} {Acta Metallurgica}\ }\textbf {\bibinfo
  {volume} {27}},\ \bibinfo {pages} {47 } (\bibinfo {year} {1979})}\BibitemShut
  {NoStop}%
\bibitem [{\citenamefont {Falk}\ and\ \citenamefont {Langer}(1998)}]{Falk98}%
  \BibitemOpen
  \bibfield  {author} {\bibinfo {author} {\bibfnamefont {M.~L.}\ \bibnamefont
  {Falk}}\ and\ \bibinfo {author} {\bibfnamefont {J.~S.}\ \bibnamefont
  {Langer}},\ }\href@noop {} {\bibfield  {journal} {\bibinfo  {journal} {Phys.\
  Rev.\ E}\ }\textbf {\bibinfo {volume} {57}},\ \bibinfo {pages} {7192}
  (\bibinfo {year} {1998})}\BibitemShut {NoStop}%
\bibitem [{\citenamefont {Tanguy}\ \emph {et~al.}(2006)\citenamefont {Tanguy},
  \citenamefont {Leonforte},\ and\ \citenamefont {Barrat}}]{Tanguy06}%
  \BibitemOpen
  \bibfield  {author} {\bibinfo {author} {\bibfnamefont {A.}~\bibnamefont
  {Tanguy}}, \bibinfo {author} {\bibfnamefont {F.}~\bibnamefont {Leonforte}}, \
  and\ \bibinfo {author} {\bibfnamefont {J.-L.}\ \bibnamefont {Barrat}},\
  }\href@noop {} {\bibfield  {journal} {\bibinfo  {journal} {The European
  Physical Journal E}\ }\textbf {\bibinfo {volume} {20}},\ \bibinfo {pages}
  {355} (\bibinfo {year} {2006})}\BibitemShut {NoStop}%
\bibitem [{\citenamefont {Schall}\ \emph {et~al.}(2007)\citenamefont {Schall},
  \citenamefont {Weitz},\ and\ \citenamefont {Spaepen}}]{Schall07}%
  \BibitemOpen
  \bibfield  {author} {\bibinfo {author} {\bibfnamefont {P.}~\bibnamefont
  {Schall}}, \bibinfo {author} {\bibfnamefont {D.~A.}\ \bibnamefont {Weitz}}, \
  and\ \bibinfo {author} {\bibfnamefont {F.}~\bibnamefont {Spaepen}},\
  }\href@noop {} {\bibfield  {journal} {\bibinfo  {journal} {Science}\ }\textbf
  {\bibinfo {volume} {318}},\ \bibinfo {pages} {1895} (\bibinfo {year}
  {2007})}\BibitemShut {NoStop}%
\bibitem [{\citenamefont {Amon}\ \emph {et~al.}(2012)\citenamefont {Amon},
  \citenamefont {Nguyen}, \citenamefont {Bruand}, \citenamefont {Crassous},\
  and\ \citenamefont {Cl\'ement}}]{Amon12}%
  \BibitemOpen
  \bibfield  {author} {\bibinfo {author} {\bibfnamefont {A.}~\bibnamefont
  {Amon}}, \bibinfo {author} {\bibfnamefont {V.}~\bibnamefont {Nguyen}},
  \bibinfo {author} {\bibfnamefont {A.}~\bibnamefont {Bruand}}, \bibinfo
  {author} {\bibfnamefont {J.}~\bibnamefont {Crassous}}, \ and\ \bibinfo
  {author} {\bibfnamefont {E.}~\bibnamefont {Cl\'ement}},\ }\href {\doibase
  10.1103/PhysRevLett.108.135502} {\bibfield  {journal} {\bibinfo  {journal}
  {Phys.\ Rev.\ Lett.}\ }\textbf {\bibinfo {volume} {108}},\ \bibinfo {pages}
  {135502} (\bibinfo {year} {2012})}\BibitemShut {NoStop}%
\bibitem [{\citenamefont {Maloney}\ and\ \citenamefont
  {Lema\^itre}(2006)}]{Maloney06a}%
  \BibitemOpen
  \bibfield  {author} {\bibinfo {author} {\bibfnamefont {C.~E.}\ \bibnamefont
  {Maloney}}\ and\ \bibinfo {author} {\bibfnamefont {A.}~\bibnamefont
  {Lema\^itre}},\ }\href@noop {} {\bibfield  {journal} {\bibinfo  {journal}
  {Phys. Rev. E}\ }\textbf {\bibinfo {volume} {74}},\ \bibinfo {pages} {016118}
  (\bibinfo {year} {2006})}\BibitemShut {NoStop}%
\bibitem [{\citenamefont {Lema\^itre}\ and\ \citenamefont
  {Caroli}(2009)}]{Lemaitre09}%
  \BibitemOpen
  \bibfield  {author} {\bibinfo {author} {\bibfnamefont {A.}~\bibnamefont
  {Lema\^itre}}\ and\ \bibinfo {author} {\bibfnamefont {C.}~\bibnamefont
  {Caroli}},\ }\href {\doibase 10.1103/PhysRevLett.103.065501} {\bibfield
  {journal} {\bibinfo  {journal} {Phys. Rev. Lett.}\ }\textbf {\bibinfo
  {volume} {103}},\ \bibinfo {pages} {065501} (\bibinfo {year}
  {2009})}\BibitemShut {NoStop}%
\bibitem [{\citenamefont {Maloney}\ and\ \citenamefont
  {Robbins}(2009)}]{Maloney09}%
  \BibitemOpen
  \bibfield  {author} {\bibinfo {author} {\bibfnamefont {C.~E.}\ \bibnamefont
  {Maloney}}\ and\ \bibinfo {author} {\bibfnamefont {M.~O.}\ \bibnamefont
  {Robbins}},\ }\href {\doibase 10.1103/PhysRevLett.102.225502} {\bibfield
  {journal} {\bibinfo  {journal} {Phys. Rev. Lett.}\ }\textbf {\bibinfo
  {volume} {102}},\ \bibinfo {pages} {225502} (\bibinfo {year}
  {2009})}\BibitemShut {NoStop}%
\bibitem [{\citenamefont {M{\"u}ller}\ and\ \citenamefont
  {Wyart}(2015)}]{Muller14}%
  \BibitemOpen
  \bibfield  {author} {\bibinfo {author} {\bibfnamefont {M.}~\bibnamefont
  {M{\"u}ller}}\ and\ \bibinfo {author} {\bibfnamefont {M.}~\bibnamefont
  {Wyart}},\ }\href {\doibase 10.1146/annurev-conmatphys-031214-014614}
  {\bibfield  {journal} {\bibinfo  {journal} {Annual Review of Condensed Matter
  Physics}\ }\textbf {\bibinfo {volume} {6}},\ \bibinfo {pages} {177} (\bibinfo
  {year} {2015})}\BibitemShut {NoStop}%
\bibitem [{\citenamefont {Lin}\ \emph {et~al.}(2015)\citenamefont {Lin},
  \citenamefont {Gueudr{\'e}}, \citenamefont {Rosso},\ and\ \citenamefont
  {Wyart}}]{Lin15}%
  \BibitemOpen
  \bibfield  {author} {\bibinfo {author} {\bibfnamefont {J.}~\bibnamefont
  {Lin}}, \bibinfo {author} {\bibfnamefont {T.}~\bibnamefont {Gueudr{\'e}}},
  \bibinfo {author} {\bibfnamefont {A.}~\bibnamefont {Rosso}}, \ and\ \bibinfo
  {author} {\bibfnamefont {M.}~\bibnamefont {Wyart}},\ }\href@noop {}
  {\bibfield  {journal} {\bibinfo  {journal} {Phys. Rev. Lett.}\ }\textbf
  {\bibinfo {volume} {115}},\ \bibinfo {pages} {168001} (\bibinfo {year}
  {2015})}\BibitemShut {NoStop}%
\bibitem [{\citenamefont {Lema{\^\i}tre}\ and\ \citenamefont
  {Caroli}(2007)}]{Lemaitre07}%
  \BibitemOpen
  \bibfield  {author} {\bibinfo {author} {\bibfnamefont {A.}~\bibnamefont
  {Lema{\^\i}tre}}\ and\ \bibinfo {author} {\bibfnamefont {C.}~\bibnamefont
  {Caroli}},\ }\href@noop {} {\bibfield  {journal} {\bibinfo  {journal} {arXiv
  preprint arXiv:0705.3122}\ } (\bibinfo {year} {2007})}\BibitemShut {NoStop}%
\bibitem [{\citenamefont {Karmakar}\ \emph {et~al.}(2010)\citenamefont
  {Karmakar}, \citenamefont {Lerner},\ and\ \citenamefont
  {Procaccia}}]{Karmakar10a}%
  \BibitemOpen
  \bibfield  {author} {\bibinfo {author} {\bibfnamefont {S.}~\bibnamefont
  {Karmakar}}, \bibinfo {author} {\bibfnamefont {E.}~\bibnamefont {Lerner}}, \
  and\ \bibinfo {author} {\bibfnamefont {I.}~\bibnamefont {Procaccia}},\ }\href
  {\doibase 10.1103/PhysRevE.82.055103} {\bibfield  {journal} {\bibinfo
  {journal} {Phys. Rev. E}\ }\textbf {\bibinfo {volume} {82}},\ \bibinfo
  {pages} {055103} (\bibinfo {year} {2010})}\BibitemShut {NoStop}%
\bibitem [{\citenamefont {Lin}\ \emph {et~al.}(2014{\natexlab{a}})\citenamefont
  {Lin}, \citenamefont {Lerner}, \citenamefont {Rosso},\ and\ \citenamefont
  {Wyart}}]{Lin14}%
  \BibitemOpen
  \bibfield  {author} {\bibinfo {author} {\bibfnamefont {J.}~\bibnamefont
  {Lin}}, \bibinfo {author} {\bibfnamefont {E.}~\bibnamefont {Lerner}},
  \bibinfo {author} {\bibfnamefont {A.}~\bibnamefont {Rosso}}, \ and\ \bibinfo
  {author} {\bibfnamefont {M.}~\bibnamefont {Wyart}},\ }\href@noop {}
  {\bibfield  {journal} {\bibinfo  {journal} {Proc.\ Natl.\ Acad.\ Sci.\
  U.S.A.}\ }\textbf {\bibinfo {volume} {111}},\ \bibinfo {pages} {14382}
  (\bibinfo {year} {2014}{\natexlab{a}})}\BibitemShut {NoStop}%
\bibitem [{\citenamefont {Lin}\ \emph {et~al.}(2014{\natexlab{b}})\citenamefont
  {Lin}, \citenamefont {Saade}, \citenamefont {Lerner}, \citenamefont {Rosso},\
  and\ \citenamefont {Wyart}}]{Lin14a}%
  \BibitemOpen
  \bibfield  {author} {\bibinfo {author} {\bibfnamefont {J.}~\bibnamefont
  {Lin}}, \bibinfo {author} {\bibfnamefont {A.}~\bibnamefont {Saade}}, \bibinfo
  {author} {\bibfnamefont {E.}~\bibnamefont {Lerner}}, \bibinfo {author}
  {\bibfnamefont {A.}~\bibnamefont {Rosso}}, \ and\ \bibinfo {author}
  {\bibfnamefont {M.}~\bibnamefont {Wyart}},\ }\href@noop {} {\bibfield
  {journal} {\bibinfo  {journal} {EPL (Europhysics Letters)}\ }\textbf
  {\bibinfo {volume} {105}},\ \bibinfo {pages} {26003} (\bibinfo {year}
  {2014}{\natexlab{b}})}\BibitemShut {NoStop}%
\bibitem [{\citenamefont {Budrikis}\ \emph {et~al.}(2017)\citenamefont
  {Budrikis}, \citenamefont {Castellanos}, \citenamefont {Sandfeld},
  \citenamefont {Zaiser},\ and\ \citenamefont {Zapperi}}]{Budrikis17}%
  \BibitemOpen
  \bibfield  {author} {\bibinfo {author} {\bibfnamefont {Z.}~\bibnamefont
  {Budrikis}}, \bibinfo {author} {\bibfnamefont {D.~F.}\ \bibnamefont
  {Castellanos}}, \bibinfo {author} {\bibfnamefont {S.}~\bibnamefont
  {Sandfeld}}, \bibinfo {author} {\bibfnamefont {M.}~\bibnamefont {Zaiser}}, \
  and\ \bibinfo {author} {\bibfnamefont {S.}~\bibnamefont {Zapperi}},\
  }\href@noop {} {\bibfield  {journal} {\bibinfo  {journal} {Nature
  communications}\ }\textbf {\bibinfo {volume} {8}},\ \bibinfo {pages} {15928}
  (\bibinfo {year} {2017})}\BibitemShut {NoStop}%
\bibitem [{\citenamefont {Lin}\ and\ \citenamefont {Wyart}(2016)}]{Lin16}%
  \BibitemOpen
  \bibfield  {author} {\bibinfo {author} {\bibfnamefont {J.}~\bibnamefont
  {Lin}}\ and\ \bibinfo {author} {\bibfnamefont {M.}~\bibnamefont {Wyart}},\
  }\href@noop {} {\bibfield  {journal} {\bibinfo  {journal} {Physical Review
  X}\ }\textbf {\bibinfo {volume} {6}},\ \bibinfo {pages} {011005} (\bibinfo
  {year} {2016})}\BibitemShut {NoStop}%
\bibitem [{\citenamefont {Hentschel}\ \emph {et~al.}(2015)\citenamefont
  {Hentschel}, \citenamefont {Jaiswal}, \citenamefont {Procaccia},\ and\
  \citenamefont {Sastry}}]{Hentschel15}%
  \BibitemOpen
  \bibfield  {author} {\bibinfo {author} {\bibfnamefont {H.}~\bibnamefont
  {Hentschel}}, \bibinfo {author} {\bibfnamefont {P.~K.}\ \bibnamefont
  {Jaiswal}}, \bibinfo {author} {\bibfnamefont {I.}~\bibnamefont {Procaccia}},
  \ and\ \bibinfo {author} {\bibfnamefont {S.}~\bibnamefont {Sastry}},\
  }\href@noop {} {\bibfield  {journal} {\bibinfo  {journal} {Physical Review
  E}\ }\textbf {\bibinfo {volume} {92}},\ \bibinfo {pages} {062302} (\bibinfo
  {year} {2015})}\BibitemShut {NoStop}%
\bibitem [{\citenamefont {Berthier}\ \emph {et~al.}(2018)\citenamefont
  {Berthier}, \citenamefont {Biroli}, \citenamefont {Ozawa}, \citenamefont
  {Tarjus},\ and\ \citenamefont {Rosso}}]{Ozawa18}%
  \BibitemOpen
  \bibfield  {author} {\bibinfo {author} {\bibfnamefont {L.}~\bibnamefont
  {Berthier}}, \bibinfo {author} {\bibfnamefont {G.}~\bibnamefont {Biroli}},
  \bibinfo {author} {\bibfnamefont {M.}~\bibnamefont {Ozawa}}, \bibinfo
  {author} {\bibfnamefont {G.}~\bibnamefont {Tarjus}}, \ and\ \bibinfo {author}
  {\bibfnamefont {A.}~\bibnamefont {Rosso}},\ }\href@noop {} {\bibfield
  {journal} {\bibinfo  {journal} {Proc.\ Natl.\ Acad.\ Sci.\ U.S.A.}\ }\textbf
  {\bibinfo {volume} {115}},\ \bibinfo {pages} {6656} (\bibinfo {year}
  {2018})}\BibitemShut {NoStop}%
\bibitem [{\citenamefont {Schober}\ and\ \citenamefont
  {Oligschleger}(1996)}]{Schober96}%
  \BibitemOpen
  \bibfield  {author} {\bibinfo {author} {\bibfnamefont {H.~R.}\ \bibnamefont
  {Schober}}\ and\ \bibinfo {author} {\bibfnamefont {C.}~\bibnamefont
  {Oligschleger}},\ }\href {\doibase 10.1103/PhysRevB.53.11469} {\bibfield
  {journal} {\bibinfo  {journal} {Phys. Rev. B}\ }\textbf {\bibinfo {volume}
  {53}},\ \bibinfo {pages} {11469} (\bibinfo {year} {1996})}\BibitemShut
  {NoStop}%
\bibitem [{\citenamefont {Ilyin}\ \emph {et~al.}(1987)\citenamefont {Ilyin},
  \citenamefont {Karpov},\ and\ \citenamefont {Parshin}}]{Ilyin87}%
  \BibitemOpen
  \bibfield  {author} {\bibinfo {author} {\bibfnamefont {M.~A.}\ \bibnamefont
  {Ilyin}}, \bibinfo {author} {\bibfnamefont {V.~G.}\ \bibnamefont {Karpov}}, \
  and\ \bibinfo {author} {\bibfnamefont {D.~A.}\ \bibnamefont {Parshin}},\
  }\href@noop {} {\bibfield  {journal} {\bibinfo  {journal} {Zh. Eksp. Teor.
  Fiz.}\ }\textbf {\bibinfo {volume} {92}},\ \bibinfo {pages} {291} (\bibinfo
  {year} {1987})}\BibitemShut {NoStop}%
\bibitem [{\citenamefont {Baity-Jesi}\ \emph {et~al.}(2015)\citenamefont
  {Baity-Jesi}, \citenamefont {Mart{\'\i}n-Mayor}, \citenamefont {Parisi},\
  and\ \citenamefont {Perez-Gaviro}}]{Baity15}%
  \BibitemOpen
  \bibfield  {author} {\bibinfo {author} {\bibfnamefont {M.}~\bibnamefont
  {Baity-Jesi}}, \bibinfo {author} {\bibfnamefont {V.}~\bibnamefont
  {Mart{\'\i}n-Mayor}}, \bibinfo {author} {\bibfnamefont {G.}~\bibnamefont
  {Parisi}}, \ and\ \bibinfo {author} {\bibfnamefont {S.}~\bibnamefont
  {Perez-Gaviro}},\ }\href@noop {} {\bibfield  {journal} {\bibinfo  {journal}
  {Physical review letters}\ }\textbf {\bibinfo {volume} {115}},\ \bibinfo
  {pages} {267205} (\bibinfo {year} {2015})}\BibitemShut {NoStop}%
\bibitem [{\citenamefont {Gartner}\ and\ \citenamefont
  {Lerner}(2016)}]{Gartner16}%
  \BibitemOpen
  \bibfield  {author} {\bibinfo {author} {\bibfnamefont {L.}~\bibnamefont
  {Gartner}}\ and\ \bibinfo {author} {\bibfnamefont {E.}~\bibnamefont
  {Lerner}},\ }\href {\doibase 10.21468/SciPostPhys.1.2.016} {\bibfield
  {journal} {\bibinfo  {journal} {SciPost Phys.}\ }\textbf {\bibinfo {volume}
  {1}},\ \bibinfo {pages} {016} (\bibinfo {year} {2016})}\BibitemShut {NoStop}%
\bibitem [{\citenamefont {Lerner}\ \emph {et~al.}(2016)\citenamefont {Lerner},
  \citenamefont {D{\"u}ring},\ and\ \citenamefont {Bouchbinder}}]{Lerner16}%
  \BibitemOpen
  \bibfield  {author} {\bibinfo {author} {\bibfnamefont {E.}~\bibnamefont
  {Lerner}}, \bibinfo {author} {\bibfnamefont {G.}~\bibnamefont {D{\"u}ring}},
  \ and\ \bibinfo {author} {\bibfnamefont {E.}~\bibnamefont {Bouchbinder}},\
  }\href@noop {} {\bibfield  {journal} {\bibinfo  {journal} {Physical Review
  Letters}\ }\textbf {\bibinfo {volume} {117}},\ \bibinfo {pages} {035501}
  (\bibinfo {year} {2016})}\BibitemShut {NoStop}%
\bibitem [{\citenamefont {Mizuno}\ \emph {et~al.}(2017)\citenamefont {Mizuno},
  \citenamefont {Shiba},\ and\ \citenamefont {Ikeda}}]{Mizuno17}%
  \BibitemOpen
  \bibfield  {author} {\bibinfo {author} {\bibfnamefont {H.}~\bibnamefont
  {Mizuno}}, \bibinfo {author} {\bibfnamefont {H.}~\bibnamefont {Shiba}}, \
  and\ \bibinfo {author} {\bibfnamefont {A.}~\bibnamefont {Ikeda}},\ }\href
  {\doibase 10.1073/pnas.1709015114} {\bibfield  {journal} {\bibinfo  {journal}
  {Proc.\ Natl.\ Acad.\ Sci.\ U.S.A.}\ }\textbf {\bibinfo {volume} {114}},\
  \bibinfo {pages} {E9767} (\bibinfo {year} {2017})}\BibitemShut {NoStop}%
\bibitem [{\citenamefont {Stanifer}\ \emph {et~al.}(2018)\citenamefont
  {Stanifer}, \citenamefont {Morse}, \citenamefont {Middleton},\ and\
  \citenamefont {Manning}}]{Stanifer18}%
  \BibitemOpen
  \bibfield  {author} {\bibinfo {author} {\bibfnamefont {E.}~\bibnamefont
  {Stanifer}}, \bibinfo {author} {\bibfnamefont {P.}~\bibnamefont {Morse}},
  \bibinfo {author} {\bibfnamefont {A.}~\bibnamefont {Middleton}}, \ and\
  \bibinfo {author} {\bibfnamefont {M.}~\bibnamefont {Manning}},\ }\href@noop
  {} {\bibfield  {journal} {\bibinfo  {journal} {arXiv preprint
  arXiv:1804.04074}\ } (\bibinfo {year} {2018})}\BibitemShut {NoStop}%
\bibitem [{\citenamefont {Wang}\ \emph {et~al.}(2018)\citenamefont {Wang},
  \citenamefont {Ninarello}, \citenamefont {Guan}, \citenamefont {Berthier},
  \citenamefont {Szamel},\ and\ \citenamefont {Flenner}}]{Wang18}%
  \BibitemOpen
  \bibfield  {author} {\bibinfo {author} {\bibfnamefont {L.}~\bibnamefont
  {Wang}}, \bibinfo {author} {\bibfnamefont {A.}~\bibnamefont {Ninarello}},
  \bibinfo {author} {\bibfnamefont {P.}~\bibnamefont {Guan}}, \bibinfo {author}
  {\bibfnamefont {L.}~\bibnamefont {Berthier}}, \bibinfo {author}
  {\bibfnamefont {G.}~\bibnamefont {Szamel}}, \ and\ \bibinfo {author}
  {\bibfnamefont {E.}~\bibnamefont {Flenner}},\ }\href@noop {} {\bibfield
  {journal} {\bibinfo  {journal} {arXiv preprint arXiv:1804.08765}\ } (\bibinfo
  {year} {2018})}\BibitemShut {NoStop}%
\bibitem [{\citenamefont {Lerner}\ and\ \citenamefont
  {Bouchbinder}(2017)}]{Lerner17}%
  \BibitemOpen
  \bibfield  {author} {\bibinfo {author} {\bibfnamefont {E.}~\bibnamefont
  {Lerner}}\ and\ \bibinfo {author} {\bibfnamefont {E.}~\bibnamefont
  {Bouchbinder}},\ }\href@noop {} {\bibfield  {journal} {\bibinfo  {journal}
  {Physical Review E}\ }\textbf {\bibinfo {volume} {96}},\ \bibinfo {pages}
  {020104} (\bibinfo {year} {2017})}\BibitemShut {NoStop}%
\bibitem [{\citenamefont {Xu}\ \emph {et~al.}(2017)\citenamefont {Xu},
  \citenamefont {Liu},\ and\ \citenamefont {Nagel}}]{Xu17}%
  \BibitemOpen
  \bibfield  {author} {\bibinfo {author} {\bibfnamefont {N.}~\bibnamefont
  {Xu}}, \bibinfo {author} {\bibfnamefont {A.~J.}\ \bibnamefont {Liu}}, \ and\
  \bibinfo {author} {\bibfnamefont {S.~R.}\ \bibnamefont {Nagel}},\ }\href
  {\doibase 10.1103/PhysRevLett.119.215502} {\bibfield  {journal} {\bibinfo
  {journal} {Phys. Rev. Lett.}\ }\textbf {\bibinfo {volume} {119}},\ \bibinfo
  {pages} {215502} (\bibinfo {year} {2017})}\BibitemShut {NoStop}%
\bibitem [{\citenamefont {Gurevich}\ \emph {et~al.}(2003)\citenamefont
  {Gurevich}, \citenamefont {Parshin},\ and\ \citenamefont
  {Schober}}]{Gurevich03}%
  \BibitemOpen
  \bibfield  {author} {\bibinfo {author} {\bibfnamefont {V.}~\bibnamefont
  {Gurevich}}, \bibinfo {author} {\bibfnamefont {D.}~\bibnamefont {Parshin}}, \
  and\ \bibinfo {author} {\bibfnamefont {H.}~\bibnamefont {Schober}},\
  }\href@noop {} {\bibfield  {journal} {\bibinfo  {journal} {Physical Review
  B}\ }\textbf {\bibinfo {volume} {67}},\ \bibinfo {pages} {094203} (\bibinfo
  {year} {2003})}\BibitemShut {NoStop}%
\bibitem [{\citenamefont {Gurarie}\ and\ \citenamefont
  {Chalker}(2003)}]{Gurarie03}%
  \BibitemOpen
  \bibfield  {author} {\bibinfo {author} {\bibfnamefont {V.}~\bibnamefont
  {Gurarie}}\ and\ \bibinfo {author} {\bibfnamefont {J.}~\bibnamefont
  {Chalker}},\ }\href@noop {} {\bibfield  {journal} {\bibinfo  {journal}
  {Physical Review B}\ }\textbf {\bibinfo {volume} {68}},\ \bibinfo {pages}
  {134207} (\bibinfo {year} {2003})}\BibitemShut {NoStop}%
\bibitem [{\citenamefont {Karpov}\ \emph {et~al.}(1983)\citenamefont {Karpov},
  \citenamefont {Klinger},\ and\ \citenamefont {Ignat’ev}}]{Karpov83}%
  \BibitemOpen
  \bibfield  {author} {\bibinfo {author} {\bibfnamefont {V.~G.}\ \bibnamefont
  {Karpov}}, \bibinfo {author} {\bibfnamefont {I.}~\bibnamefont {Klinger}}, \
  and\ \bibinfo {author} {\bibfnamefont {F.~N.}\ \bibnamefont {Ignat’ev}},\
  }\href@noop {} {\bibfield  {journal} {\bibinfo  {journal} {Zh. Eksp. Teor.
  Fiz.}\ }\textbf {\bibinfo {volume} {84}},\ \bibinfo {pages} {760} (\bibinfo
  {year} {1983})}\BibitemShut {NoStop}%
\bibitem [{\citenamefont {Karpov}\ and\ \citenamefont
  {Parshin}(1985)}]{Karpov85}%
  \BibitemOpen
  \bibfield  {author} {\bibinfo {author} {\bibfnamefont {V.}~\bibnamefont
  {Karpov}}\ and\ \bibinfo {author} {\bibfnamefont {D.}~\bibnamefont
  {Parshin}},\ }\href {http://jetp.ac.ru/cgi-bin/dn/e{\_}061{\_}06{\_}1308.pdf}
  {\bibfield  {journal} {\bibinfo  {journal} {Soviet physics, JETP}\ }\textbf
  {\bibinfo {volume} {61}},\ \bibinfo {pages} {1308} (\bibinfo {year}
  {1985})}\BibitemShut {NoStop}%
\bibitem [{\citenamefont {Picard}\ \emph {et~al.}(2004)\citenamefont {Picard},
  \citenamefont {Ajdari}, \citenamefont {Lequeux},\ and\ \citenamefont
  {Bocquet}}]{Picard04}%
  \BibitemOpen
  \bibfield  {author} {\bibinfo {author} {\bibfnamefont {G.}~\bibnamefont
  {Picard}}, \bibinfo {author} {\bibfnamefont {A.}~\bibnamefont {Ajdari}},
  \bibinfo {author} {\bibfnamefont {F.}~\bibnamefont {Lequeux}}, \ and\
  \bibinfo {author} {\bibfnamefont {L.}~\bibnamefont {Bocquet}},\ }\href
  {\doibase 10.1140/epje/i2004-10054-8} {\bibfield  {journal} {\bibinfo
  {journal} {The European Physical Journal E}\ }\textbf {\bibinfo {volume}
  {15}},\ \bibinfo {pages} {371} (\bibinfo {year} {2004})}\BibitemShut
  {NoStop}%
\bibitem [{\citenamefont {Baret}\ \emph {et~al.}(2002)\citenamefont {Baret},
  \citenamefont {Vandembroucq},\ and\ \citenamefont {Roux}}]{Baret02}%
  \BibitemOpen
  \bibfield  {author} {\bibinfo {author} {\bibfnamefont {J.-C.}\ \bibnamefont
  {Baret}}, \bibinfo {author} {\bibfnamefont {D.}~\bibnamefont {Vandembroucq}},
  \ and\ \bibinfo {author} {\bibfnamefont {S.}~\bibnamefont {Roux}},\ }\href
  {\doibase 10.1103/PhysRevLett.89.195506} {\bibfield  {journal} {\bibinfo
  {journal} {Phys. Rev. Lett.}\ }\textbf {\bibinfo {volume} {89}},\ \bibinfo
  {pages} {195506} (\bibinfo {year} {2002})}\BibitemShut {NoStop}%
\bibitem [{\citenamefont {Nicolas}\ \emph {et~al.}(2017)\citenamefont
  {Nicolas}, \citenamefont {Ferrero}, \citenamefont {Martens},\ and\
  \citenamefont {Barrat}}]{Nicolas17}%
  \BibitemOpen
  \bibfield  {author} {\bibinfo {author} {\bibfnamefont {A.}~\bibnamefont
  {Nicolas}}, \bibinfo {author} {\bibfnamefont {E.~E.}\ \bibnamefont
  {Ferrero}}, \bibinfo {author} {\bibfnamefont {K.}~\bibnamefont {Martens}}, \
  and\ \bibinfo {author} {\bibfnamefont {J.-L.}\ \bibnamefont {Barrat}},\
  }\href@noop {} {\bibfield  {journal} {\bibinfo  {journal} {arXiv preprint
  arXiv:1708.09194}\ } (\bibinfo {year} {2017})}\BibitemShut {NoStop}%
\bibitem [{Note1()}]{Note1}%
  \BibitemOpen
  \bibinfo {note} {Blocks are also inserted at $\lambda = \lambda _{\protect
  \text {min}}$ as discussed earlier. However, as long as a finite fraction of
  blocks are reinserted they give a dominant contribution to the density of
  states. For example, in a mean-field approximation blocks perform L\'{e}vy
  Flight of index $\mu = 1$ \cite {Lin16} in which diffusion and drift are
  comparable so that a finite fraction of blocks fails in opposite direction of
  the drift and is reinserted.}\BibitemShut {Stop}%
\bibitem [{\citenamefont {{M. P. Allen and D. J. Tildesley}}(1991)}]{Allen91}%
  \BibitemOpen
  \bibfield  {author} {\bibinfo {author} {\bibnamefont {{M. P. Allen and D. J.
  Tildesley}}},\ }\href@noop {} {\emph {\bibinfo {title} {{Computer Simulations
  of Liquids}}}}\ (\bibinfo  {publisher} {Oxford University Press, London},\
  \bibinfo {year} {1991})\BibitemShut {NoStop}%
\bibitem [{\citenamefont {Efros}\ and\ \citenamefont
  {Shklovskii}(1975)}]{Efros75}%
  \BibitemOpen
  \bibfield  {author} {\bibinfo {author} {\bibfnamefont {A.~L.}\ \bibnamefont
  {Efros}}\ and\ \bibinfo {author} {\bibfnamefont {B.~I.}\ \bibnamefont
  {Shklovskii}},\ }\href {http://stacks.iop.org/0022-3719/8/i=4/a=003}
  {\bibfield  {journal} {\bibinfo  {journal} {Journal of Physics C: Solid State
  Physics}\ }\textbf {\bibinfo {volume} {8}},\ \bibinfo {pages} {L49} (\bibinfo
  {year} {1975})}\BibitemShut {NoStop}%
\bibitem [{\citenamefont {Kapteijns}\ \emph {et~al.}(2018)\citenamefont
  {Kapteijns}, \citenamefont {Ji}, \citenamefont {Brito}, \citenamefont
  {Wyart},\ and\ \citenamefont {Lerner}}]{Kapteijns18}%
  \BibitemOpen
  \bibfield  {author} {\bibinfo {author} {\bibfnamefont {G.}~\bibnamefont
  {Kapteijns}}, \bibinfo {author} {\bibfnamefont {W.}~\bibnamefont {Ji}},
  \bibinfo {author} {\bibfnamefont {C.}~\bibnamefont {Brito}}, \bibinfo
  {author} {\bibfnamefont {M.}~\bibnamefont {Wyart}}, \ and\ \bibinfo {author}
  {\bibfnamefont {E.}~\bibnamefont {Lerner}},\ }\href@noop {} {\bibfield
  {journal} {\bibinfo  {journal} {arXiv preprint arXiv:1808.00018}\ } (\bibinfo
  {year} {2018})}\BibitemShut {NoStop}%
\end{thebibliography}%


%

\appendix

\section{Potential shape dynamics}\label{App.A}
Here we derive Eq.~\eqref{eq:3} of the main text. Following Eqs.~(\ref{eq:1},\ref{eq:2}), the tilted potential under shear stress in a block $i$ is
\begin{align}
\tilde{u}_i(s,\; \delta \sigma_i)
=\frac{1}{2!}\lambda_{i}s^{2}+\frac{1}{3!}\kappa_{i}s^{3}+\frac{1}{4!}s^{4}-s\delta \sigma_i.
\end{align}
which can be equivalently expressed as
\begin{align}
 \tilde{u}_i(s,\; \delta \sigma_i)=
\frac{1}{2!}\tilde{\lambda}_{i}\left(s-s_{0}\right)^{2}+\frac{1}{3!}\tilde{\kappa}_{i}\left(s-s_{0}\right)^{3}+\frac{1}{4!}\left(s-s_{0}\right)^{4},
\label{eq:A2}
\end{align}
where $s_0$ is the displacement that corresponds to the new minimum, and $(\tilde{\lambda}_{i}, \tilde{\kappa}_{i})$ are the new Taylor expansion coefficients around $s= s_0$. The relation between $(\tilde{\lambda}_{i}, \tilde{\kappa}_{i}, s_0)$ and $(\lambda_i, \kappa_i, \delta \sigma_i)$ reads:
\begin{equation}
\begin{cases}
\displaystyle
\delta\kappa_{i}\equiv \tilde{\kappa}_{i}-\kappa_{i}=s_{0}
\\
\displaystyle
\delta\lambda_{i}\equiv \tilde{\lambda}_{i}-\lambda_{i}=\tilde{\kappa}_{i}s_{0}-\frac{1}{2}s_{0}^{2}
\\
\displaystyle
\delta \sigma_i=\tilde{\lambda}_{i}s_{0}-\frac{1}{2}\tilde{\kappa}_{i}s_{0}^{2}+\frac{1}{6}s_{0}^{3}

\end{cases}
.
 \label{eq:A3}
\end{equation}
In the limit $\delta \sigma_i \to 0$ we obtain
\begin{equation}
\begin{cases}
\displaystyle
\frac{\partial \kappa_{i}}{\partial \sigma_i}\equiv\lim_{\delta \sigma_i\rightarrow0}\frac{\delta\kappa_{i}}{\delta \sigma_i}=\frac{1}{\lambda_{i}}
\vspace*{.7em}
\\
\displaystyle
\frac{\partial\lambda_{i}}{\partial \sigma_i}\equiv\lim_{\delta \sigma_i\rightarrow0}\frac{\delta\lambda_{i}}{\delta \sigma_i}=\frac{\kappa_{i}}{\lambda_{i}}
\end{cases}
.
 \label{eq:A4}
\end{equation}

\section{Derivation of the parabola $\kappa^2-8\lambda/3=0$ after the block fails.} \label{App.B}
When the block fails, it drops to a new minimum.  The potential expands at the new minimum is:
\begin{align}
u(s)
=\frac{1}{2!} \lambda s^{2}+\frac{1}{3!}\kappa s^{3}+\frac{1}{4!}s^{4}.
\end{align}
At the  inflection point, it satisfies:
\begin{align}
\begin{cases}
\displaystyle
\frac{du(s)}{ds}=0\\
\displaystyle
\frac{d^{2}u(s)}{ds^{2}}=0
\end{cases}
\end{align}
We eliminate $s$ and obtain the relation: $\kappa^2-8 \lambda/3=0$.

\section{Derivation of the Jacobian $|\partial(\Delta\sigma, \Phi)/\partial(\lambda, \kappa)|$} \label{App.C}
To calculate the Jacobian $|\partial(\Delta\sigma, \Phi)/\partial(\lambda, \kappa)|$ we first express $\sigma$ as a function of $\lambda$ and $\kappa$. To this end we substitute $\lambda=(\kappa^{2}-\Phi)/2$ (Eq.~\eqref{eq:5}) into $\partial\kappa/\partial \Delta\sigma = 1/\lambda$ (Eq.~\eqref{eq:3}):
\begin{equation}
\left(\frac{\kappa^{2}-\Phi}{2}\right)\frac{\partial\kappa}{\partial \Delta\sigma}=1 ,
 \label{eq:B1}
\end{equation}
so that
\begin{align}
\frac{1}{6}\frac{\partial\left(\kappa^{3}-3\Phi\kappa\right)}{\partial \Delta\sigma} & =1 ,
 \label{eq:B2}
\end{align}
and therefore
\begin{equation}
\Delta\sigma=\frac{\kappa^{3}-3\lambda\kappa}{3}+c_{0}.
 \label{eq:B3}
\end{equation}
where $c_{0}$ is an integration constant. Hence, the Jacobian is:
\begin{equation}
\left\Vert\frac{\partial(\Delta\sigma, \Phi)}{\partial(\lambda, \kappa)}\right\Vert= \left|\frac{\partial \Delta\sigma}{\partial\lambda}\frac{\partial\Phi}{\partial\kappa}-\frac{\partial \Delta\sigma}{\partial\kappa}\frac{\partial\Phi}{\partial\lambda}\right|=2\lambda .
 \label{eq:B5}
\end{equation}

\section{Distance to the instability}\label{App.D}
When $\Phi> 0 $ we choose the integration constant $c_0$ in Eq.\ \eqref{eq:B3} such that $x=0$ at $\lambda=0$. We find that
\begin{equation}
x=\frac{\lambda\kappa-\left(\kappa -  \sign{(\kappa)}\sqrt{\kappa^{2}-2\lambda}\right)\left(\kappa^{2}-2\lambda\right)}{3} .
\label{eq:C1}
\end{equation}
In the limit $\lambda\ll\kappa^{2}$
\begin{equation}
x=\frac{1}{2}\frac{\lambda^{2}}{\kappa}.
\label{eq:C2}
\end{equation}

When a block fails, it is reinserted on the parabola $\kappa^2 - 8/3 \lambda= 0$ and therefore the distance to instability of the reinserted block is
\begin{align}
  x^{*}= \frac{1}{12}\kappa^3 = \frac{2}{3}\Phi^{3/2}.
\label{eq:C3}
\end{align}

\section{System preparation protocol} \label{App.E}
The glass system that we use is identical to the one by [32]. It consists of a binary mixture of point-masses (`particles'). All details including all parameters values can
be found in its supplementary material.

We consider glasses obtained by two different system preparation protocols. Each glass is represented by an ensemble of one thousand independent realisations. Each realisation is obtained by a temperature quench that starts from a state of thermal equilibrium at a temperature $T = T_0$ that is higher than the glass transition temperature. The two quenching protocols are: (\emph{i}) an instantaneous quench,  referred to as `rapid quench', in which fully overdamped dynamics are used until all the particles' velocities have converged to zero. Here  $T_{0}=\frac{\varepsilon}{k_{B}}$, where $\varepsilon$ is a microscopic energy scale and $k_{B}$ is Boltzmann's constant.  (\emph{ii}) a continuous quench, referred to as `slow quench', in which the system is first solidified at a cooling rate of $\dot{T}=10^{-3}T_0/t_c$ until the temperature $T=0.1T_0$ is reached;
overdamped dynamics are  then employed to remove the remaining heat. Here, $t_c\equiv\sqrt{md^{2}/\varepsilon}$, $m$ is the mass of each particle and $d$ is the diameter of the `small' particles.

\section{Measurement of $\gamma_{\min}$} \label{App.F}

\begin{figure}[htb]
  \includegraphics[width=.95\linewidth]{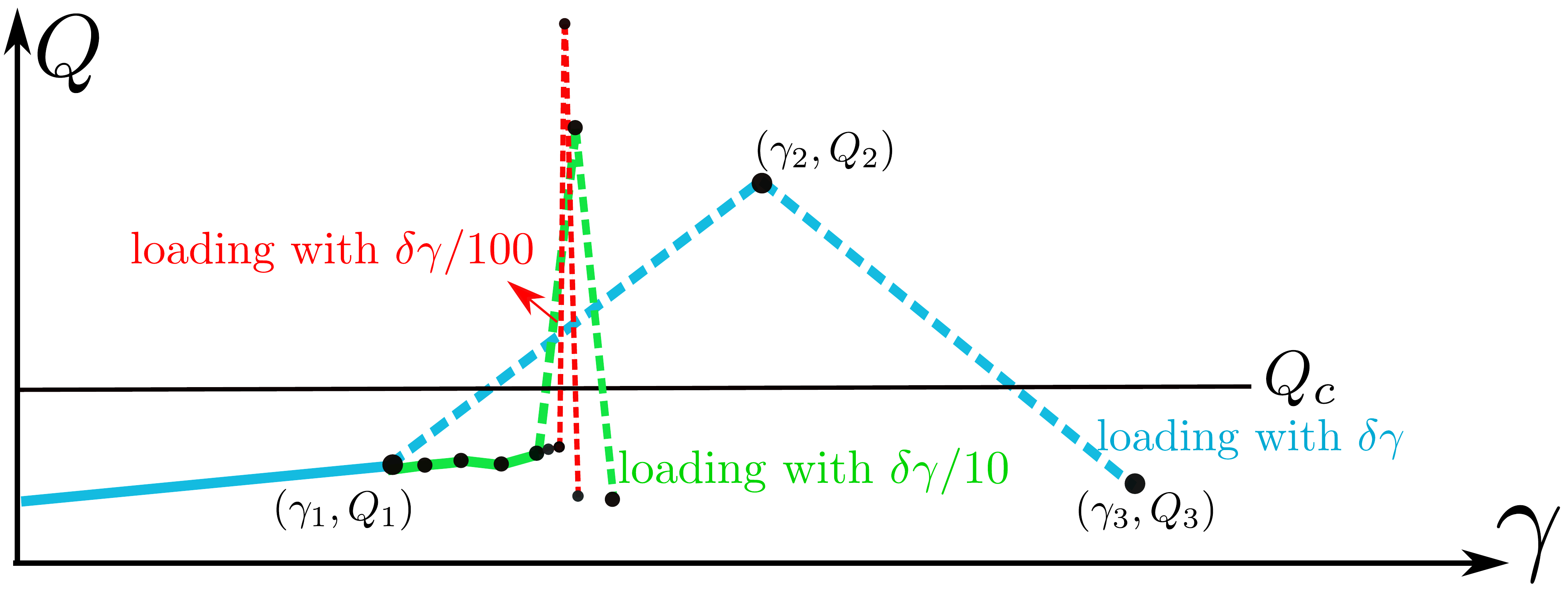}
  \caption{Sketch of the protocol to detect the instability. The cyan curve, the blue curve and the red curve correspond to shear steps $\delta\gamma$, $\delta\gamma/10$, $\delta\gamma/100$, respectively. Dashed curves represent the trial steps. If at some point $Q_2> \max{\left(Q_1, Q_3, Q_c\right)}$ trial steps are reversed and the strain step is reduced by a factor 10.}
  \label{measure}
\end{figure}

To accurately measure $\gamma_{\min}$ a loading protocol has been developed in which the applied shear $\gamma$ is adaptively refined when an instability. More practically, when an instability is detected the system is rewound to its last known equilibrium state before the instability. The instability using smaller steps for $\delta \gamma$. This protocol is repeated a number of times, such that the value of $\gamma$ at which the instability occurs is characterised with a sufficient accuracy. The instability is detected by monitoring a quantity $Q \equiv \max \left| \delta \pvec{r}_{i}' \right| / ( L \delta\gamma )$ where $L$ is the linear system size and $\max\,|\delta \pvec{r}_{i}'|$ is the maximal change in the non-affine displacement of a particle, for the given increment in applied affine shear $\delta\gamma$. When the response is elastic, $Q$ is of order one. (Note that the factor $1/L$, used to define a dimensionless $Q$, changes with the system size. We do not expect that this size dependence affects our results because $L\sim N^\frac{1}{3}$ changes by a factor of $3.3$ between $N=2000$ and $N=64000$, which is less than the typical fluctuation of $Q$ in different realisations at fixed $N$.). However, if there is a shear transformation, $Q$ is significantly higher.  To detect the shear transformation we set a threshold $Q_c$ to be much larger than the typical $Q$ in elastic shearing  and then monitor the three successive $Q_1$, $Q_2$, $Q_3$ that results from the strain history $\gamma_1<\gamma_2<\gamma_3$. If $Q_2>Q_c$ and  $Q_1, Q_3<Q_2$,  we go back from $\gamma_3$ to $\gamma_1$ and set the strain increment $\delta\gamma/10$ (see Fig.\ \ref{measure}). This is repeated until the strain increment is smaller than $10^{-6}$. At this final resolution, we use an additional condition $100Q_3<Q_2$ to make sure that $Q$ is discontinuous which implies that the instability is present. In our simulations, the initial strain increment $\delta\gamma=10^{-4}$ and $Q_c=100$. In the end, we could ensure the error of $\gamma_{\min}$ to be less than $10^{-5}$.

\section{Finite size effects in $D_L(\omega)$ and relation to $P(\gamma_{\min})$} \label{App.H}
\begin{figure}[htb]
  \centering
  \includegraphics[width=.95\linewidth]{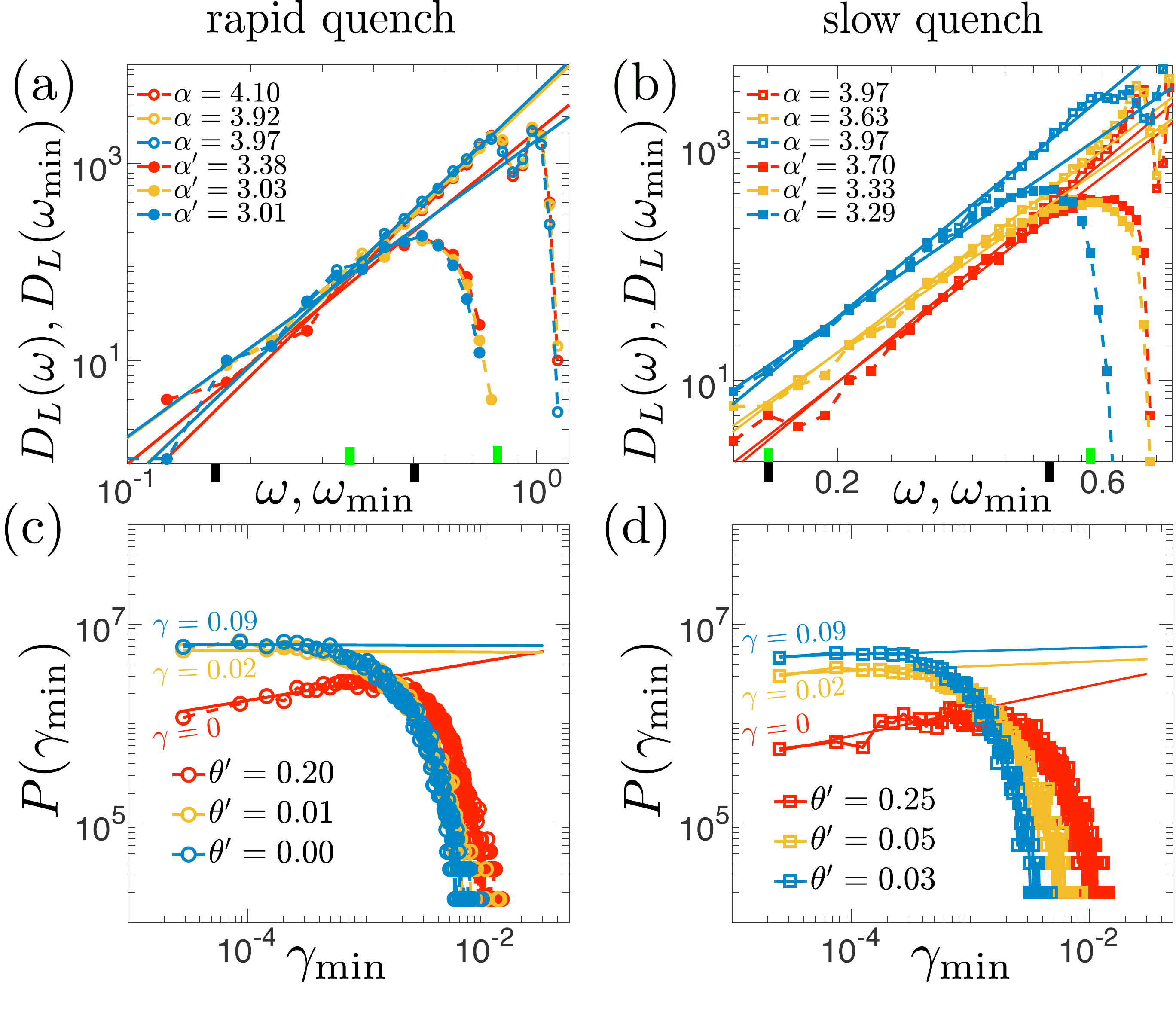}
  \caption{The left column shows results obtained after a rapid quench of a system of $N=16000$ particles and the right column after a slow quench for $N=32000$. (a,b) The comparison between $D_L(\omega)$ and $D_L(\omega_{\min})$ at three representative $\gamma=0$, $0.02$, $0.09$ (red, yellow, blue). The black markers on the axes indicate the fitting range of $D_L(\omega_{\min})$.  The green markers on the axes indicate the same the fitting range used in the main text (cf.\ Figs.~\ref{fig2}(g,h)). (c,d) The distribution of $P(\gamma_{\min})$ at the same three representative $\gamma=0$, $0.02$, $0.09$ (red, yellow, blue). }
  \label{alpha'}
\end{figure}

We now discuss finite size effects that affect the distribution $D_L(\omega)$ at low frequencies. They stem from the fact that the states of the glass are sampled at a given value of the accumulated strain $\gamma$ (see the inset of Fig.~\ref{fig2}(a) in the main text). As a consequence, there is a finite probability density to sample a state that is arbitrarily close to the next instability and therefore we are more likely to sample a shear transformation. This sampling also prevents us to measure $\theta$ by directly fitting $P(\gamma_{\min})$. Namely, as there will be a uniform probability density of finding a shear transformation close to the instability, $P(\gamma_{\min})$ will be uniform at small values of $\gamma_{\min}$. Note that $\gamma= 0$ is a special point because the system is always sampled directly after an avalanche.

\begin{figure}[htb]
  \centering
  \includegraphics[width=.95\linewidth]{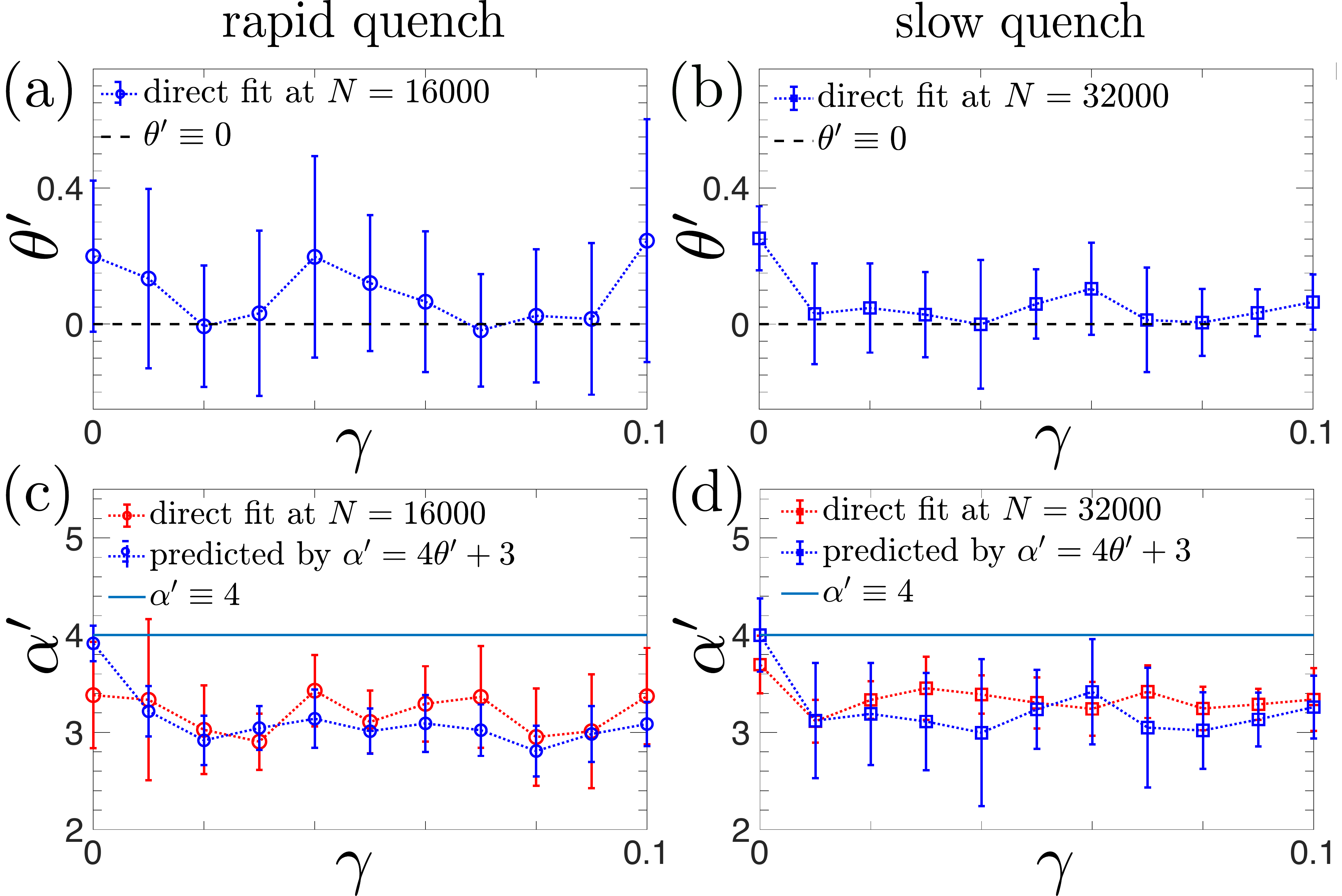}
  \caption{The left column shows results obtained after a rapid quench for $N=16000$ using 1000 realisations and the right column results obtained after a slow quench for $N=32000$ using 5000 realisations. (a,b) $\theta'$ is obtained by fitting $P(\gamma_{\min})$ at different $\gamma$. (c,d) Measured values of $\alpha'$ (red line) are in good agreement with the theoretical prediction $\alpha'= 4\theta' + 3$ (blue line).}
  \label{alpha'_theta'}
\end{figure}

To confirm this explanation we measured exponents $\theta'$ and $\alpha'$ defined by $P(\gamma_{\min})\sim \gamma_{\min}^{\theta'}$ and $D_L(\omega_{\min})\sim \omega_{\min}^{\alpha'}$, respectively, where $\omega_{\min}$ is the frequency of the softest quasi-localised mode. Figs.\ \ref{alpha'}(a,b) compare distributions $D_L(\omega)$ and $D_L(\omega_{\min})$ at three representative strain $\gamma=0$, $0.02$, $0.09$. As observed, $\alpha$ and $\alpha'$ are clearly different. The distributions $P(\gamma_{\min})$ are also shown in Figs.\ \ref{alpha'}(c,d) at the  three representative strain. Also the values of $\theta'$ are clearly different from $\theta$, as displayed in main text in Figs.~\ref{fig3}(a,b).

In Figs.\ \ref{alpha'_theta'}(a,b) we report values of $\theta'$ respectively in rapidly and slowly quenched glasses. In both cases $\theta'$ is practically $0$. This confirms our explanation presented above.
The corresponding values of $\alpha'$ are shown in Figs.\ \ref{alpha'_theta'}(c,d), for which a  reasonable agreement is found with the values predicted by the theory. This confirms our explanation: when a small $\omega_{\min}$ is measured it almost always corresponds to a system sampled by chance close to the instability. Therefore, $D_L(\omega_{\min})$ is dominated by shear transformations even if $D_L(\omega)$ is not. This can be clearly seen by comparing regions where $\alpha= 4$ for all $\gamma$ (both shear transformations and passive modes contribute significantly to $D_L(\omega)$) but $\alpha' < 4$ (shear transformations are dominant).

\section{Comparison of $\alpha$ obtained using different fitting ranges of $\omega$} \label{App.G}
\begin{figure}[htb]
  \centering
  \includegraphics[width=.95\linewidth]{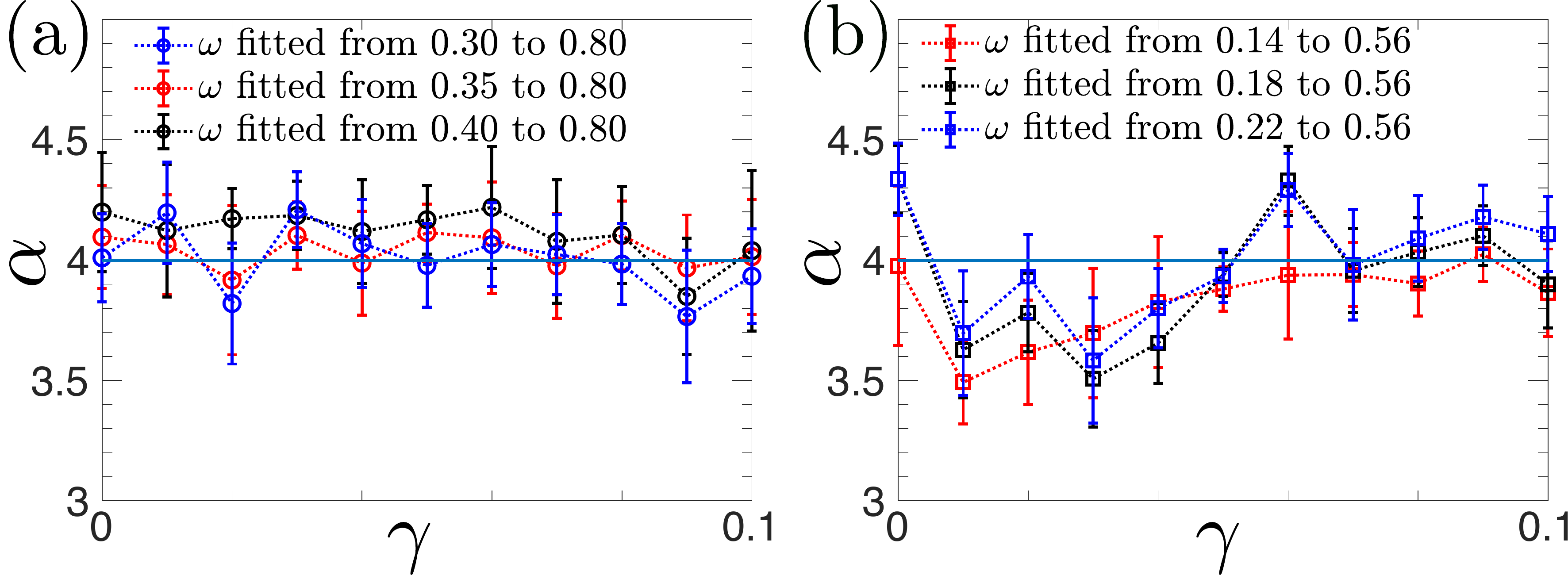}
  \caption{$\alpha$ measured at different lower bounds of fitting ranges after (a) a rapid quench and (b) a slow quench. The results that are reported in the main text (Figs.~\ref{fig3}(c,d)) correspond to the red lines.}
  \label{ranges}
\end{figure}

To test the robustness of the fit of the exponent $\alpha$, we measure it using different lower bounds of the range over which $D_L(\omega)$ is fit. Note that at this point plane waves have been filtered out by the protocol described in the main text. This has set the upper bound of the fitting range to be there where the power-law scaling is clearly interrupted by the plane waves.  In Fig.\ \ref{ranges}(a,b) we show that there is a range of lower bounds for which the measured values of $\alpha$ are robust in rapidly and slowly quenched glasses, respectively.

\section{$\theta$ and $\alpha$ in steady state in rapidly quenched systems } \label{App.I}
\begin{figure}[htb]
  \centering
  \includegraphics[width=.95\linewidth]{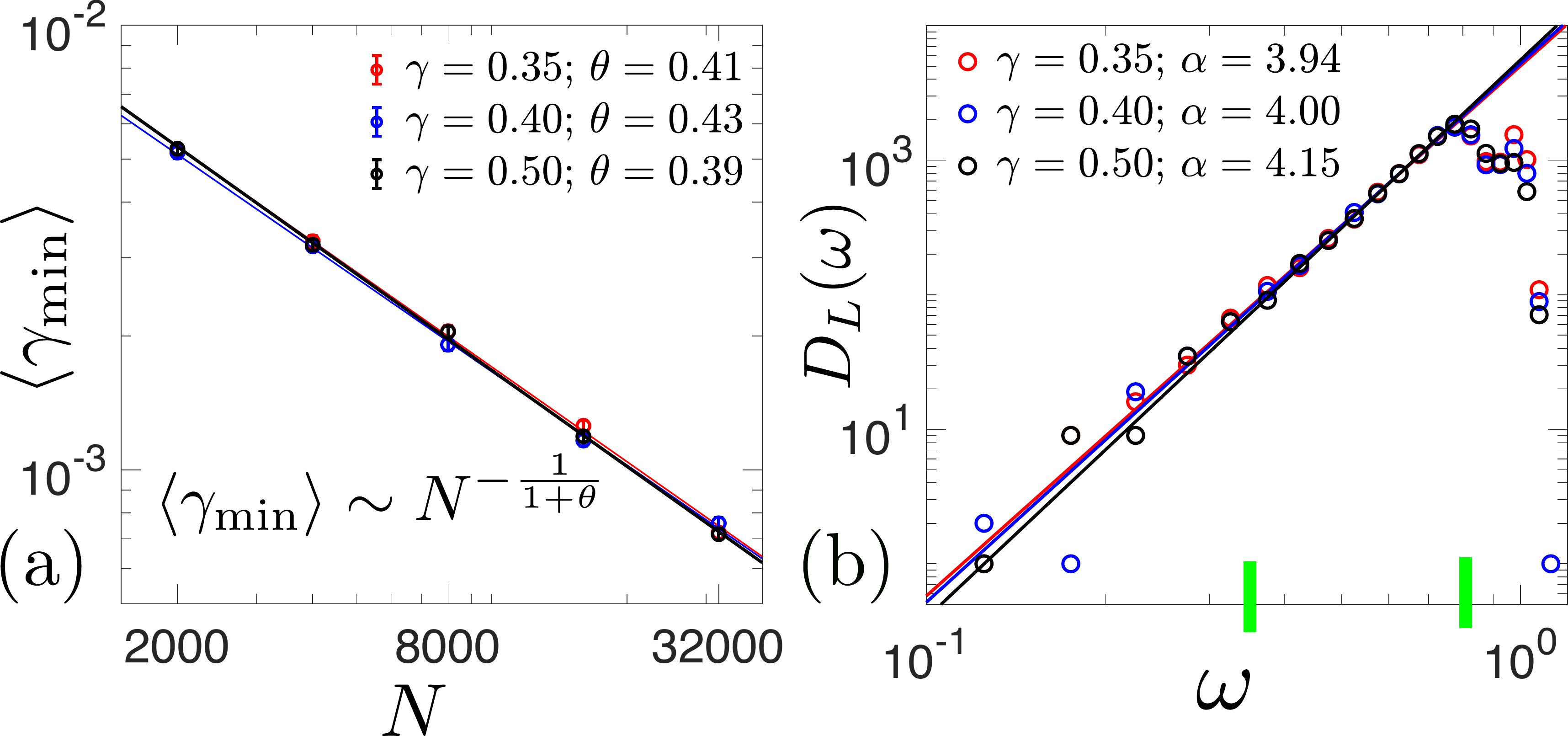}
  \caption{(a) $\theta$  obtained from finite size scaling analysis in steady state. (b) The density of states of quasi-localised vibrational modes $D_L(\omega)$ at low frequencies $\omega$  for $N=16000$ after modes with a participation ratio above the threshold $e_{c}$ have been removed. The green markers on the axes indicate the same fitting range that is used in the main text (cf.\ Fig.~\ref{fig2}(g)).}
  \label{alpha_steady}
\end{figure}

In the steady state, $\theta$ and $\alpha$ should no longer depend on the system preparation. In Figs.\ \ref{alpha_steady}(a,b) we show $\theta$ and $\alpha$ at large strains for the system prepared by a rapid quench. Clearly, $\theta$ converges to a constant larger than $1/4$ and $\alpha\simeq4$.

\end{document}